\documentclass[%
 aip,
 amsmath,amssymb,
 reprint,%
]{revtex4-1}

\usepackage{graphicx}% Include figure files
\usepackage{dcolumn}% Align table columns on decimal point
\usepackage{bm}% bold math
\usepackage[utf8]{inputenc}
\usepackage[T1]{fontenc}
\usepackage{mathptmx}
\usepackage{etoolbox}
\usepackage{siunitx}
\usepackage{color}
\usepackage{braket}

\makeatletter
\def\@email#1#2{%
 \endgroup
 \patchcmd{\titleblock@produce}
  {\frontmatter@RRAPformat}
  {\frontmatter@RRAPformat{\produce@RRAP{*#1\href{mailto:#2}{#2}}}\frontmatter@RRAPformat}
  {}{}
}%
\makeatother
\begin{document}

\preprint{AIP/123-QED}

\title[]{Diamond quantum sensors in microfluidics technology}
% Force line breaks with \\
\author{Masazumi Fujiwara}
 % \altaffiliation[Guest scholar: ]{Institute of Physics, Humboldt University of Berlin}%Lines break automatically or can be forced with \\
  \email{masazumi@okayama-u.ac.jp}
% \author{Keisuke Oshimi}%
\affiliation{ 
Department of Chemistry, Graduate School of Environmental, Life, Natural Science and Technology, 
Okayama University, 3-1-1, Tsushimanaka, Kita-ku, Okayama-shi, Okayama 700-8530, Japan
%\\This line break forced with \textbackslash\textbackslash
}%

% \author{C. Author}
%  \homepage{http://www.Second.institution.edu/~Charlie.Author.}
% \affiliation{%
% Second institution and/or address%\\This line break forced% with \\
% }%

\date{\today}% It is always \today, today,
             %  but any date may be explicitly specified

\begin{abstract}
Diamond quantum sensing is an emerging technology for probing multiple physico-chemical parameters in the nano- to micro-scale dimensions within diverse chemical and biological contexts. 
Integrating these sensors into microfluidic devices enables the precise quantification and analysis of small sample volumes in microscale channels. 
In this Perspective, we present recent advancements in the integration of diamond quantum sensors with microfluidic devices and explore their prospects with a focus on forthcoming technological developments.
\end{abstract}

\maketitle

% \begin{quotation}
% The ``lead paragraph'' is encapsulated with the \LaTeX\ 
% \verb+quotation+ environment and is formatted as a single paragraph before the first section heading. 
% (The \verb+quotation+ environment reverts to its usual meaning after the first sectioning command.) 
% Note that numbered references are allowed in the lead paragraph.
% %
% The lead paragraph will only be found in an article being prepared for the journal \textit{Chaos}.
% \end{quotation}

\section*{Introduction}
Microfluidics holds promise for the development of devices that provide rapid, efficient, and sensitive analysis of chemical and biological samples~\cite{ngo2014dna,MAEKI202280,rothbauer2018recent}. The miniaturization of analytical systems into chip devices significantly reduces sample numbers and volumes~\cite{WenAnalChem2008,HattoriAnalChem2019}. 
Additionally, analytes can be purified and concentrated using nanostructures fabricated in microfluids, which enables low-noise and high-sensitivity detection~\cite{WenAnalChem2008,HattoriAnalChem2019}. 
The concept of microfluidics extends to engineering miniature biological tissues inside channels, thus mimicking physiological conditions using organ-on-chip technology~\cite{leung2022guide,VollertsenBiomicro2021}. 
However, implementing small, sensitive, and multimodal sensors into micrometer-sized channels poses a challenge; therefore, diamond quantum sensors are attractive for overcoming this challenge in microfluidic technologies.

\begin{figure*}[th!]
 \centering
 \includegraphics[width=17cm]{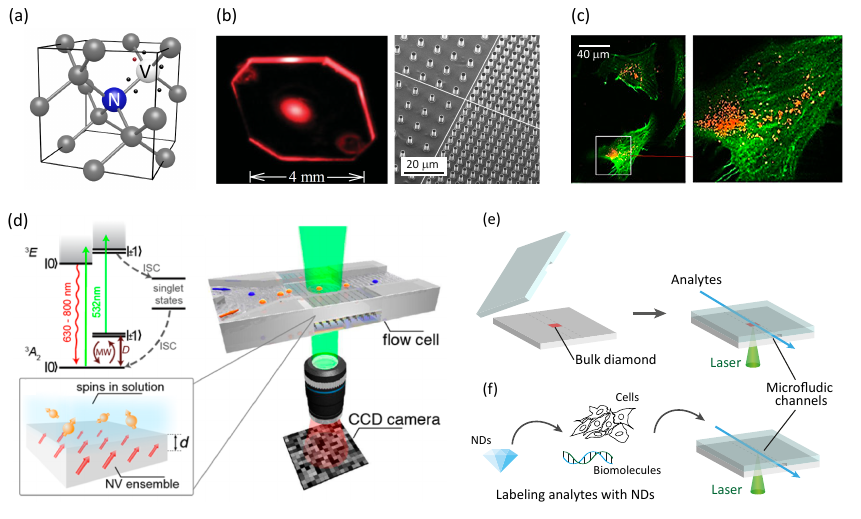}
 \caption{Graphical concept of using diamond NV quantum sensors with microfluidics. 
 {
 (a) Schematic of the NV structure. 
 (b) Left: a photograph of bulk diamond chip with red NV fluorescence. 
 Figure reproduced with permission from Phys. Rev. B \textbf{85}, 121202 (2012)~\cite{PhysRevB.85.121202}. Copyright 2012 American Physical Society.
 Right: an electron microscope image showing NV-embedded nanopillar arrays (Losero et al., Sci. Rep. \textbf{13}, 5909 (2023)~\cite{losero2023neuronal}; licensed under a Creative Commons Attribution (CC BY 4.0) license.)
 (c) Fluorescence microscope images capturing fluorescent NDs in HeLa cells. Red: NDs. Green: pHrodo Green Dextran (Nie et al., Nanomaterials \textbf{11}, 1837  (2021)~\cite{nano11071837}; licensed under a Creative Commons Attribution (CC BY 4.0) license.)
 (d) Schematic of simplified energy level structure of NV centers and microfluidic spin detection with freely diffusing spins above the bulk diamond sensors. $\ket{0}$ and $\ket{\pm1}$ are the spin sublevels for $m_S = 0$ and $m_S = \pm 1$, respectively.
 MW: microwave. $D$: zero-field splitting of the spin sublevels. 
 ISC: intersystem crossing. 
 $d$: distance between the surface and the NV spin layer.
 Figure reproduced with slight modifications with permission from Nano Lett. \textbf{13}, 4093 (2013)~\cite{ziem2013highly}. Copyright 2013 American Chemical Society. 
 (e, f) Two types of microfluidics with diamond quantum sensors using bulk chips and NDs.
 }
 }
 \label{figure 1}
\end{figure*}

Diamond quantum sensors enable the ultrasensitive and multimodal analyses of various chemical substances and biological samples, including biomolecules~\cite{ferrier2009microwave, ziem2013highly, wackerlig2016applications, rendler2017optical, miller2020spin, haziza2017fluorescent}, cells, and small organisms~\cite{kucsko2013nanometre, simpson2017non, toraille2018optical, claveau2018fluorescent, nie2021quantum,davis2018mapping, van2020evaluation, fujiwara2020real}. 
The inert nature of the diamond surface facilitates intact and nontoxic sample analysis. 
The sensing multimodality stems from the dependence of the resonance frequencies or relaxation times of electron spins in nitrogen vacancy (NV) centers on the magnetic field~\cite{rondin2014magnetometry, maclaurin2013nanoscale, horowitz2012electron}, electric field~\cite{dolde2011electric, iwasaki2017direct, bian2021nanoscale}, or temperature~\cite{neumann2013high, fujiwara2021diamond, wang2018magnetic}.
{NV centers, a major type of color defect centers in diamond  (Fig.~\ref{figure 1}a), create their energy states deeply embedded in the wide bandgap of diamond, emitting deep-red fluorescence in the range of 630--800 nm upon green optical excitation.
These NV centers can be artificially generated in both bulk diamonds and nanodiamonds (NDs), as depicted in Fig.~\ref{figure 1}b, c.
As illustrated in Fig.~\ref{figure 1}d, the electronic ground state of NV centers are spin triplet (${^3A_2}$) with
electron spin sublevels of $m_S = 0$ and $m_S = \pm 1$ .
The NV spin state can be initialized to $m_S = 0$, under optical excitation, as the electrons in $m_S = \pm 1$ of the excited state (${^3E}$) undergo intersystem crossing to $m_S = 0$ of the ground state via the metastable singlet states. 
Upon application of a microwave resonant to the transition of $m_S = 0 \to \pm1$ in the ground state (typically at $\sim 2.87$ GHz), the fluorescence intensity is decreased, a phenomenon known as optically detected magnetic resonance (ODMR).
}
Recent studies have expanded measurement targets to include parameters such as pH~\cite{rendler2017optical, fujisaku2019ph} and radical molecules~\cite{nie2021quantum, Nie-NanoLett-2022} by exploiting the sensitivity of NV spin relaxation times to magnetic spin noise.

In microfluidic channels, NV centers are placed near the samples, either several nanometers below the surface of bulk diamonds or in NDs. 
The ODMR of samples flowing in the microfluidic channel is measured using NV layers near the surface of the bulk diamonds~\cite{rendler2017optical, ziem2013highly, allert2022microfluidic} {(Fig.~\ref{figure 1}d, e)}. 
Alternatively, NDs are introduced into the sample, and the resultant ND-labeled sample is fed into the microfluidic channel for ODMR detection~\cite{oshimi2022glass} {(Fig.~\ref{figure 1}f)}.
The scaling down of ODMR-based analysis into microfluidic channels necessitates the miniaturization and optimization of fluorescence collection and microwave excitation, as ODMR relies on the double resonance of light and microwaves, requiring technological development for more efficient devices. 
This Perspective highlights recent advances in the implementation of diamond-based ODMR technologies in microfluidic channels.

\section*{Utilization of Microfluidics in Diamond Quantum Sensing}
The unique sensing capabilities of diamond quantum sensors, particularly their small volume and near-surface sensitivity, have led to their integration with microfluidic technologies since the early stages of development. 
In one approach, bulk diamond chips containing NV ensembles on the surface were bonded with polydimethylsiloxane (PDMS)-based microfluidic systems to detect ferritin molecules and gadolinium spin labels in solutions and cells~\cite{ziem2013highly, rendler2017optical} (Fig.~\ref{figure 1}a).
Microfluidics offers precise manipulation of small analyte volumes on the order of picoliters within submillimeter spaces~\cite{Andrich-NanoLett-2014, Lim-NanoLett-2015}, facilitating efficient interfacing of analytes with diamond quantum sensors and maximizing their sensing capabilities. 
For instance, NV-based diamond quantum sensors integrated into microfluidic channels have demonstrated two-dimensional nuclear magnetic resonance (NMR) spectroscopy with a detection volume of approximately 40 pL, aided by a spin-prepolarization magnet~\cite{Smits-SciAdv-2019}.
Nanostructures such as nanopillars and mesopores can be introduced to enhance the surface sensitivity of diamond quantum sensors. 
Liu et al. demonstrated that metal--organic-framework (MOF) mesoporous structures on a diamond surface increased the effective surface area and boosted NV-NMR sensitivity~\cite{Liu-MOF-NanoLett-2022,PhysRevX.10.021053}.
Additionally, interfacing spin-active analytes with diamond nanostructures can potentially polarize 13C-labeled molecules, such as acetic anhydride and pyruvate~\cite{doi:10.1073/pnas.0810190106,doi:10.1021/acs.jpclett.2c03785,PhysRevResearch.4.043179}, and 
{has been studied to further increase NMR sensitivity~\cite{PhysRevB.103.014434}.}
% can be subsequently used for the hyperpolarization of analytes to further increase NMR sensitivity~\cite{alvarez2015local,doi:10.1021/jacs.2c13830}.

In addition to PDMS-based microfluidics, diamond quantum sensors have been utilized in various microfluidic channels. 
An example approach involves the direct fabrication of microfluidic channels inside bulk diamonds to enhance the interfacing with analytes and surrounding diamond surfaces~\cite{PICOLLO2017193}. 
Although the fabrication of NV layers around these channels has yet to be achieved, ongoing efforts have employed femtosecond laser writing for this purpose~\cite{https://doi.org/10.1002/qute.201900006, Jedrkiewicz:17}.
Furthermore, NDs have been applied in diverse microfluidic channels. 
For instance, Miller et al. employed antibody-functionalized ND quantum sensors on a paper-fluidic platform to create lateral flow assay test kits for HIV-1 RNA, which exhibited exceptional sensitivity down to zepto-molar levels~\cite{miller2020spin}. 
Additionally, glass-capillary microfluids have been utilized in conjunction with NDs for the optical manipulation and sorting of nanomaterials, with prospects of ND-quantum sensing for certain analytes~\cite{doi:10.1021/acsanm.0c00274}.

\begin{figure*}[th!]
 \centering
 \includegraphics[width=17cm]{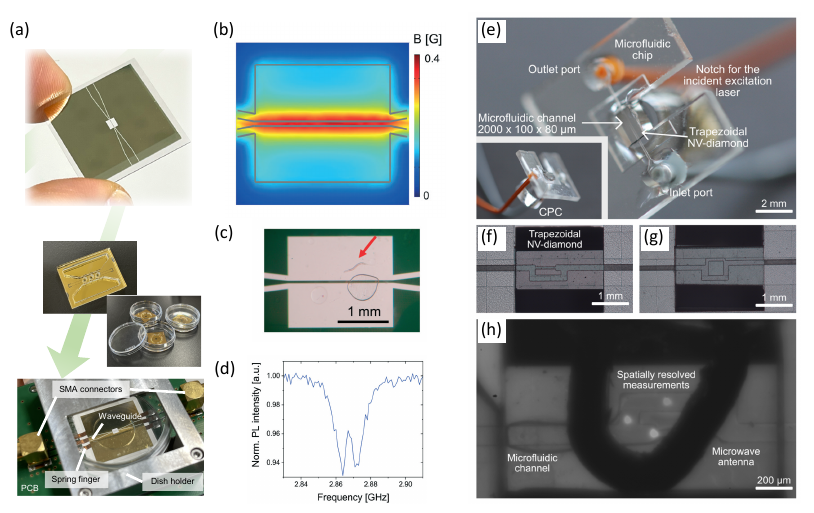}
 \caption{Recent advancement in implementing NV-ODMR technologies into on-chip devices. (a) Photograph of glass-patternable notch-shaped microwave antenna and its assembling with biological assay platforms including multi-well plates and glass bottom dishes. These devices are docked with an electric circuit board for ODMR measurements. (b) Corresponding microwave magnetic field intensity distribution map in the notch area. (c, d) Nematode worm placed in the notch area and the corresponding ODMR spectrum observed in the worm. 
 Figures reproduced with slight modifications with permission from Lab Chip. \textbf{22}, 2519 (2022)~\cite{oshimi2022glass}. Copyright 2022 Royal Society of Chemistry. 
 (e) Photograph of a microfluidic quantum sensing platform for a  straight channel. The NV center-doped top side ($2.0 \times 1.0$ mm functions as a sensor and forms the bottom of the channel. (f, g) Example assembly with arbitrary channel geometries and dimensions to demonstrate the platform's versatility.
 (h) In operando bright-field microscopy image of the complete assembly inside a microscale NV experiment. The coaxial microwave antenna appears as a dark loop positioned on the microfluidic chip. 
 Figures reproduced with slight modifications with permission from Lab Chip. \textbf{22}, 4831 (2022)~\cite{allert2022microfluidic}. Copyright 2022 Royal Society of Chemistry.
 }
 \label{figure 2}
\end{figure*}

\section*{Technical Implementation of ODMR in Microfluidics}
As mentioned earlier, optimizing microwave excitation and fluorescence collection is crucial for ODMR measurements. 
Recent advancements have significantly improved the technological components.
Traditionally, diamond quantum sensing has often employed simple coils or thin wires as antennas for microwave excitation owing to their simplicity and ease of handling~\cite{fujiwara2021diamond}. 
However, these antennas suffer from a significant impedance mismatch with conventional cables or circuitry designed for 50 \si{\ohm}, resulting in a decreased spin excitation efficiency or necessitating high-power amplifiers, even for millimeter-sized uniform excitation.
To address this, Oshimi et al. developed a notch-shaped coplanar microwave waveguide antenna on a glass plate designed specifically for on-chip ODMR detection of the fluorescence of NDs (Fig.~\ref{figure 2}a--d).
A notch area with a size of $1.5\times2.0 \ \si{\mm}^2$ introduced in a 50-\si{\ohm} coplanar waveguide ensured uniform microwave excitation across the notch area and exhibited gigahertz broadband characteristics with a reflection loss less than 10\%~\cite{oshimi2022glass}. 
Notably, the ODMR signal intensity in the detection area can be quantitatively predicted using numerical simulations based on the calculated magnetic field intensity of microwaves via optical Bloch equations~\cite{dreau2011avoiding}. 
This device successfully demonstrated the designed ODMR signal intensity over the detection area for various biological samples including cells, tissues, and worms. 
This study highlights the potential and significance of the computational design of microwave architectures for the future integration of diamond quantum sensing with microfluidic approaches.

For optical excitation and fluorescence collection in diamond quantum sensing, a microscope objective is preferred because of its high compatibility with microfluidics technology. 
Although simple optical excitation by beams focusing on NVs is common in the existing literature, efforts are underway to fabricate optical waveguides within bulk diamonds to exploit the high refractive index of diamonds ($n=2.4$).
This high index of diamond compared with that of water ($n=1.33$) or air ($n=1.0$) presents challenges for the efficient optical excitation and fluorescence collection of NVs through the microscope objective, particularly when air or liquid media are present between the NV and objective. 
Adopting a waveguide approach offers enhanced efficiency for optical excitation and fluorescence collection in microfluidics, as demonstrated by bulk-chip diamond microscopes and nanophotonic diamond quantum devices~\cite{clevenson2015broadband}.
Eaton et al. described various approaches for fabricating photonic waveguides with NV-sensing components using laser-writing technology~\cite{https://doi.org/10.1002/qute.201900006}. 
Laser writing not only enables the creation of photonic waveguides in diamond but also generates NV centers~\cite{chen2017laser} in combination with annealing, potentially facilitating efficient all-in-one processing of microfluidic diamond quantum sensors.

In addition to the technical requirements associated with diamond quantum sensing, it is imperative to consider the prerequisites inherent in lab-on-a-chip applications. 
Allert et al. have highlighted the following criteria: (i) the adaptability and optical transparency of the diamond device structure to accommodate diverse microfluidic operations, and (ii) the device's biocompatibility~\cite{allert2022microfluidic}. 
With these considerations in mind, they have introduced a fully integrated microfluidic platform for bulk-based diamond quantum sensing, successfully implementing various NV center-based sensing modalities for chemical analysis. This includes the detection of paramagnetic ions and the realization of high-resolution microscale NV-NMR (Fig.~\ref{figure 2}e--h).

\section*{Outlook}
In this Perspective, I have summarized the efforts to integrate diamond quantum sensors with microfluidic technologies. 
Diamond quantum sensor technology is highly compatible with microfluidics and provides high sensitivity with multimodal measurements in limited microscale areas in fluidic channels. 
In the early stages of the development of diamond quantum sensors, microfluidic channels were employed to efficiently interface analytes with quantum-sensing detection areas.
However, recent research has focused on the development of a more effective integration with microfluidics to deploy high-sensitivity and multimodal sensing capabilities in chemical analyses and biological assays.
It is anticipated that these efforts will extend to several key applications of microfluidics including organ-on-chip, wearable, and high-throughput assay devices. 
For these applications, the following aspects must be considered: 
(1) {Multimodal quantum sensing, currently at a functional stage~\cite{10.1063/5.0031502,
PhysRevApplied.17.014009,10.3389/frqst.2023.1220015}, needs to be optimized for microfluidic applications, with a focus on specific measurement targets.
}
(2) Sample concentration methods on the surface of diamonds must be considered because the sensitivity of quantum sensing is scaled to the diamond surface area, whereas the number of molecules (analytes) is scaled to the volume. 
(3) Optical and microwave access to diamond quantum sensors in wearable devices must be developed. 
(4) Various microfluidic materials must be employed to improve the culture of organ-on-chip devices.
Finally, following the success of diamond quantum sensors, other solid-state quantum sensors have been developed, including those based on silicon carbide~\cite{castelletto2020silicon,doi:10.1126/sciadv.adg2080} and hexagonal boron nitride~\cite{gottscholl2021spin,doi:10.1080/23746149.2023.2206049}.
Quantum sensors provide microfluidics technology with a useful toolbox for microscale sensing.

\section*{Acknowledgements}
MF acknowledges funding from JSPS-KAKENHI (20H00335, 20KK0317), NEDO (JPNP20004), AMED (JP23zf0127004), JST (JPMJMI21G1), the RSK Sanyo Foundation, and the Asahi Glass Foundation.

\section*{Data availability statement}
No new data were created or analysed in this study.

% \nocite{*}
% \bibliography{rsc}% Produces the bibliography via BibTeX.

\begin{thebibliography}{60}%
\makeatletter
\providecommand \@ifxundefined [1]{%
 \@ifx{#1\undefined}
}%
\providecommand \@ifnum [1]{%
 \ifnum #1\expandafter \@firstoftwo
 \else \expandafter \@secondoftwo
 \fi
}%
\providecommand \@ifx [1]{%
 \ifx #1\expandafter \@firstoftwo
 \else \expandafter \@secondoftwo
 \fi
}%
\providecommand \natexlab [1]{#1}%
\providecommand \enquote  [1]{``#1''}%
\providecommand \bibnamefont  [1]{#1}%
\providecommand \bibfnamefont [1]{#1}%
\providecommand \citenamefont [1]{#1}%
\providecommand \href@noop [0]{\@secondoftwo}%
\providecommand \href [0]{\begingroup \@sanitize@url \@href}%
\providecommand \@href[1]{\@@startlink{#1}\@@href}%
\providecommand \@@href[1]{\endgroup#1\@@endlink}%
\providecommand \@sanitize@url [0]{\catcode `\\12\catcode `\$12\catcode
  `\&12\catcode `\#12\catcode `\^12\catcode `\_12\catcode `\%12\relax}%
\providecommand \@@startlink[1]{}%
\providecommand \@@endlink[0]{}%
\providecommand \url  [0]{\begingroup\@sanitize@url \@url }%
\providecommand \@url [1]{\endgroup\@href {#1}{\urlprefix }}%
\providecommand \urlprefix  [0]{URL }%
\providecommand \Eprint [0]{\href }%
\providecommand \doibase [0]{http://dx.doi.org/}%
\providecommand \selectlanguage [0]{\@gobble}%
\providecommand \bibinfo  [0]{\@secondoftwo}%
\providecommand \bibfield  [0]{\@secondoftwo}%
\providecommand \translation [1]{[#1]}%
\providecommand \BibitemOpen [0]{}%
\providecommand \bibitemStop [0]{}%
\providecommand \bibitemNoStop [0]{.\EOS\space}%
\providecommand \EOS [0]{\spacefactor3000\relax}%
\providecommand \BibitemShut  [1]{\csname bibitem#1\endcsname}%
\let\auto@bib@innerbib\@empty
%</preamble>
\bibitem [{\citenamefont {Ngo}\ \emph {et~al.}(2014)\citenamefont {Ngo},
  \citenamefont {Wang}, \citenamefont {Fales}, \citenamefont {Nicholson},
  \citenamefont {Woods},\ and\ \citenamefont {Vo-Dinh}}]{ngo2014dna}%
  \BibitemOpen
  \bibfield  {author} {\bibinfo {author} {\bibfnamefont {H.~T.}\ \bibnamefont
  {Ngo}}, \bibinfo {author} {\bibfnamefont {H.-N.}\ \bibnamefont {Wang}},
  \bibinfo {author} {\bibfnamefont {A.~M.}\ \bibnamefont {Fales}}, \bibinfo
  {author} {\bibfnamefont {B.~P.}\ \bibnamefont {Nicholson}}, \bibinfo {author}
  {\bibfnamefont {C.~W.}\ \bibnamefont {Woods}}, \ and\ \bibinfo {author}
  {\bibfnamefont {T.}~\bibnamefont {Vo-Dinh}},\ }\bibfield  {title} {\enquote
  {\bibinfo {title} {Dna bioassay-on-chip using sers detection for dengue
  diagnosis},}\ }\href@noop {} {\bibfield  {journal} {\bibinfo  {journal}
  {Analyst}\ }\textbf {\bibinfo {volume} {139}},\ \bibinfo {pages} {5655--5659}
  (\bibinfo {year} {2014})}\BibitemShut {NoStop}%
\bibitem [{\citenamefont {Maeki}\ \emph {et~al.}(2022)\citenamefont {Maeki},
  \citenamefont {Uno}, \citenamefont {Niwa}, \citenamefont {Okada},\ and\
  \citenamefont {Tokeshi}}]{MAEKI202280}%
  \BibitemOpen
  \bibfield  {author} {\bibinfo {author} {\bibfnamefont {M.}~\bibnamefont
  {Maeki}}, \bibinfo {author} {\bibfnamefont {S.}~\bibnamefont {Uno}}, \bibinfo
  {author} {\bibfnamefont {A.}~\bibnamefont {Niwa}}, \bibinfo {author}
  {\bibfnamefont {Y.}~\bibnamefont {Okada}}, \ and\ \bibinfo {author}
  {\bibfnamefont {M.}~\bibnamefont {Tokeshi}},\ }\bibfield  {title} {\enquote
  {\bibinfo {title} {Microfluidic technologies and devices for lipid
  nanoparticle-based rna delivery},}\ }\href {\doibase
  https://doi.org/10.1016/j.jconrel.2022.02.017} {\bibfield  {journal}
  {\bibinfo  {journal} {Journal of Controlled Release}\ }\textbf {\bibinfo
  {volume} {344}},\ \bibinfo {pages} {80--96} (\bibinfo {year}
  {2022})}\BibitemShut {NoStop}%
\bibitem [{\citenamefont {Rothbauer}, \citenamefont {Zirath},\ and\
  \citenamefont {Ertl}(2018)}]{rothbauer2018recent}%
  \BibitemOpen
  \bibfield  {author} {\bibinfo {author} {\bibfnamefont {M.}~\bibnamefont
  {Rothbauer}}, \bibinfo {author} {\bibfnamefont {H.}~\bibnamefont {Zirath}}, \
  and\ \bibinfo {author} {\bibfnamefont {P.}~\bibnamefont {Ertl}},\ }\bibfield
  {title} {\enquote {\bibinfo {title} {Recent advances in microfluidic
  technologies for cell-to-cell interaction studies},}\ }\href@noop {}
  {\bibfield  {journal} {\bibinfo  {journal} {Lab Chip}\ }\textbf {\bibinfo
  {volume} {18}},\ \bibinfo {pages} {249--270} (\bibinfo {year}
  {2018})}\BibitemShut {NoStop}%
\bibitem [{\citenamefont {Wen}\ \emph {et~al.}(2008)\citenamefont {Wen},
  \citenamefont {Legendre}, \citenamefont {Bienvenue},\ and\ \citenamefont
  {Landers}}]{WenAnalChem2008}%
  \BibitemOpen
  \bibfield  {author} {\bibinfo {author} {\bibfnamefont {J.}~\bibnamefont
  {Wen}}, \bibinfo {author} {\bibfnamefont {L.~A.}\ \bibnamefont {Legendre}},
  \bibinfo {author} {\bibfnamefont {J.~M.}\ \bibnamefont {Bienvenue}}, \ and\
  \bibinfo {author} {\bibfnamefont {J.~P.}\ \bibnamefont {Landers}},\
  }\bibfield  {title} {\enquote {\bibinfo {title} {Purification of nucleic
  acids in microfluidic devices},}\ }\href {\doibase 10.1021/ac8014998}
  {\bibfield  {journal} {\bibinfo  {journal} {Analytical Chemistry}\ }\textbf
  {\bibinfo {volume} {80}},\ \bibinfo {pages} {6472--6479} (\bibinfo {year}
  {2008})},\ \bibinfo {note} {pMID: 18754652},\ \Eprint
  {http://arxiv.org/abs/https://doi.org/10.1021/ac8014998}
  {https://doi.org/10.1021/ac8014998} \BibitemShut {NoStop}%
\bibitem [{\citenamefont {Hattori}\ \emph {et~al.}(2019)\citenamefont
  {Hattori}, \citenamefont {Shimada}, \citenamefont {Yasui}, \citenamefont
  {Kaji},\ and\ \citenamefont {Baba}}]{HattoriAnalChem2019}%
  \BibitemOpen
  \bibfield  {author} {\bibinfo {author} {\bibfnamefont {Y.}~\bibnamefont
  {Hattori}}, \bibinfo {author} {\bibfnamefont {T.}~\bibnamefont {Shimada}},
  \bibinfo {author} {\bibfnamefont {T.}~\bibnamefont {Yasui}}, \bibinfo
  {author} {\bibfnamefont {N.}~\bibnamefont {Kaji}}, \ and\ \bibinfo {author}
  {\bibfnamefont {Y.}~\bibnamefont {Baba}},\ }\bibfield  {title} {\enquote
  {\bibinfo {title} {Micro- and nanopillar chips for continuous separation of
  extracellular vesicles},}\ }\href {\doibase 10.1021/acs.analchem.8b05538}
  {\bibfield  {journal} {\bibinfo  {journal} {Analytical Chemistry}\ }\textbf
  {\bibinfo {volume} {91}},\ \bibinfo {pages} {6514--6521} (\bibinfo {year}
  {2019})},\ \bibinfo {note} {pMID: 31035752},\ \Eprint
  {http://arxiv.org/abs/https://doi.org/10.1021/acs.analchem.8b05538}
  {https://doi.org/10.1021/acs.analchem.8b05538} \BibitemShut {NoStop}%
\bibitem [{\citenamefont {Leung}\ \emph {et~al.}(2022)\citenamefont {Leung},
  \citenamefont {De~Haan}, \citenamefont {Ronaldson-Bouchard}, \citenamefont
  {Kim}, \citenamefont {Ko}, \citenamefont {Rho}, \citenamefont {Chen},
  \citenamefont {Habibovic}, \citenamefont {Jeon}, \citenamefont {Takayama}
  \emph {et~al.}}]{leung2022guide}%
  \BibitemOpen
  \bibfield  {author} {\bibinfo {author} {\bibfnamefont {C.~M.}\ \bibnamefont
  {Leung}}, \bibinfo {author} {\bibfnamefont {P.}~\bibnamefont {De~Haan}},
  \bibinfo {author} {\bibfnamefont {K.}~\bibnamefont {Ronaldson-Bouchard}},
  \bibinfo {author} {\bibfnamefont {G.-A.}\ \bibnamefont {Kim}}, \bibinfo
  {author} {\bibfnamefont {J.}~\bibnamefont {Ko}}, \bibinfo {author}
  {\bibfnamefont {H.~S.}\ \bibnamefont {Rho}}, \bibinfo {author} {\bibfnamefont
  {Z.}~\bibnamefont {Chen}}, \bibinfo {author} {\bibfnamefont {P.}~\bibnamefont
  {Habibovic}}, \bibinfo {author} {\bibfnamefont {N.~L.}\ \bibnamefont {Jeon}},
  \bibinfo {author} {\bibfnamefont {S.}~\bibnamefont {Takayama}},  \emph
  {et~al.},\ }\bibfield  {title} {\enquote {\bibinfo {title} {A guide to the
  organ-on-a-chip},}\ }\href@noop {} {\bibfield  {journal} {\bibinfo  {journal}
  {Nature Reviews Methods Primers}\ }\textbf {\bibinfo {volume} {2}},\ \bibinfo
  {pages} {33} (\bibinfo {year} {2022})}\BibitemShut {NoStop}%
\bibitem [{\citenamefont {Vollertsen}\ \emph {et~al.}(2021)\citenamefont
  {Vollertsen}, \citenamefont {Vivas}, \citenamefont {van Meer}, \citenamefont
  {van~den Berg}, \citenamefont {Odijk},\ and\ \citenamefont {van~der
  Meer}}]{VollertsenBiomicro2021}%
  \BibitemOpen
  \bibfield  {author} {\bibinfo {author} {\bibfnamefont {A.~R.}\ \bibnamefont
  {Vollertsen}}, \bibinfo {author} {\bibfnamefont {A.}~\bibnamefont {Vivas}},
  \bibinfo {author} {\bibfnamefont {B.}~\bibnamefont {van Meer}}, \bibinfo
  {author} {\bibfnamefont {A.}~\bibnamefont {van~den Berg}}, \bibinfo {author}
  {\bibfnamefont {M.}~\bibnamefont {Odijk}}, \ and\ \bibinfo {author}
  {\bibfnamefont {A.~D.}\ \bibnamefont {van~der Meer}},\ }\bibfield  {title}
  {\enquote {\bibinfo {title} {{Facilitating implementation of organs-on-chips
  by open platform technology}},}\ }\href {\doibase 10.1063/5.0063428}
  {\bibfield  {journal} {\bibinfo  {journal} {Biomicrofluidics}\ }\textbf
  {\bibinfo {volume} {15}},\ \bibinfo {pages} {051301} (\bibinfo {year}
  {2021})},\ \Eprint
  {http://arxiv.org/abs/https://pubs.aip.org/aip/bmf/article-pdf/doi/10.1063/5.0063428/14597190/051301\_1\_online.pdf}
  {https://pubs.aip.org/aip/bmf/article-pdf/doi/10.1063/5.0063428/14597190/051301\_1\_online.pdf}
  \BibitemShut {NoStop}%
\bibitem [{\citenamefont {Ferrier}\ \emph {et~al.}(2009)\citenamefont
  {Ferrier}, \citenamefont {Romanuik}, \citenamefont {Thomson}, \citenamefont
  {Bridges},\ and\ \citenamefont {Freeman}}]{ferrier2009microwave}%
  \BibitemOpen
  \bibfield  {author} {\bibinfo {author} {\bibfnamefont {G.~A.}\ \bibnamefont
  {Ferrier}}, \bibinfo {author} {\bibfnamefont {S.~F.}\ \bibnamefont
  {Romanuik}}, \bibinfo {author} {\bibfnamefont {D.~J.}\ \bibnamefont
  {Thomson}}, \bibinfo {author} {\bibfnamefont {G.~E.}\ \bibnamefont
  {Bridges}}, \ and\ \bibinfo {author} {\bibfnamefont {M.~R.}\ \bibnamefont
  {Freeman}},\ }\bibfield  {title} {\enquote {\bibinfo {title} {A microwave
  interferometric system for simultaneous actuation and detection of single
  biological cells},}\ }\href@noop {} {\bibfield  {journal} {\bibinfo
  {journal} {Lab Chip}\ }\textbf {\bibinfo {volume} {9}},\ \bibinfo {pages}
  {3406--3412} (\bibinfo {year} {2009})}\BibitemShut {NoStop}%
\bibitem [{\citenamefont {Ziem}\ \emph {et~al.}(2013)\citenamefont {Ziem},
  \citenamefont {Götz}, \citenamefont {Zappe}, \citenamefont {Steinert},\ and\
  \citenamefont {Wrachtrup}}]{ziem2013highly}%
  \BibitemOpen
  \bibfield  {author} {\bibinfo {author} {\bibfnamefont {F.~C.}\ \bibnamefont
  {Ziem}}, \bibinfo {author} {\bibfnamefont {N.~S.}\ \bibnamefont {Götz}},
  \bibinfo {author} {\bibfnamefont {A.}~\bibnamefont {Zappe}}, \bibinfo
  {author} {\bibfnamefont {S.}~\bibnamefont {Steinert}}, \ and\ \bibinfo
  {author} {\bibfnamefont {J.}~\bibnamefont {Wrachtrup}},\ }\bibfield  {title}
  {\enquote {\bibinfo {title} {Highly sensitive detection of physiological
  spins in a microfluidic device},}\ }\href@noop {} {\bibfield  {journal}
  {\bibinfo  {journal} {Nano Lett.}\ }\textbf {\bibinfo {volume} {13}},\
  \bibinfo {pages} {4093--4098} (\bibinfo {year} {2013})}\BibitemShut {NoStop}%
\bibitem [{\citenamefont {Wackerlig}\ and\ \citenamefont
  {Schirhagl}(2016)}]{wackerlig2016applications}%
  \BibitemOpen
  \bibfield  {author} {\bibinfo {author} {\bibfnamefont {J.}~\bibnamefont
  {Wackerlig}}\ and\ \bibinfo {author} {\bibfnamefont {R.}~\bibnamefont
  {Schirhagl}},\ }\bibfield  {title} {\enquote {\bibinfo {title} {Applications
  of molecularly imprinted polymer nanoparticles and their advances toward
  industrial use: a review},}\ }\href@noop {} {\bibfield  {journal} {\bibinfo
  {journal} {Anal. Chem.}\ }\textbf {\bibinfo {volume} {88}},\ \bibinfo {pages}
  {250--261} (\bibinfo {year} {2016})}\BibitemShut {NoStop}%
\bibitem [{\citenamefont {Rendler}\ \emph {et~al.}(2017)\citenamefont
  {Rendler}, \citenamefont {Neburkova}, \citenamefont {Zemek}, \citenamefont
  {Kotek}, \citenamefont {Zappe}, \citenamefont {Chu}, \citenamefont {Cigler},\
  and\ \citenamefont {Wrachtrup}}]{rendler2017optical}%
  \BibitemOpen
  \bibfield  {author} {\bibinfo {author} {\bibfnamefont {T.}~\bibnamefont
  {Rendler}}, \bibinfo {author} {\bibfnamefont {J.}~\bibnamefont {Neburkova}},
  \bibinfo {author} {\bibfnamefont {O.}~\bibnamefont {Zemek}}, \bibinfo
  {author} {\bibfnamefont {J.}~\bibnamefont {Kotek}}, \bibinfo {author}
  {\bibfnamefont {A.}~\bibnamefont {Zappe}}, \bibinfo {author} {\bibfnamefont
  {Z.}~\bibnamefont {Chu}}, \bibinfo {author} {\bibfnamefont {P.}~\bibnamefont
  {Cigler}}, \ and\ \bibinfo {author} {\bibfnamefont {J.}~\bibnamefont
  {Wrachtrup}},\ }\bibfield  {title} {\enquote {\bibinfo {title} {Optical
  imaging of localized chemical events using programmable diamond quantum
  nanosensors},}\ }\href@noop {} {\bibfield  {journal} {\bibinfo  {journal}
  {Nat. Commun.}\ }\textbf {\bibinfo {volume} {8}},\ \bibinfo {pages} {1--9}
  (\bibinfo {year} {2017})}\BibitemShut {NoStop}%
\bibitem [{\citenamefont {Miller}\ \emph {et~al.}(2020)\citenamefont {Miller},
  \citenamefont {Bezinge}, \citenamefont {Gliddon}, \citenamefont {Huang},
  \citenamefont {Dold}, \citenamefont {Gray}, \citenamefont {Heaney},
  \citenamefont {Dobson}, \citenamefont {Nastouli}, \citenamefont {Morton}
  \emph {et~al.}}]{miller2020spin}%
  \BibitemOpen
  \bibfield  {author} {\bibinfo {author} {\bibfnamefont {B.~S.}\ \bibnamefont
  {Miller}}, \bibinfo {author} {\bibfnamefont {L.}~\bibnamefont {Bezinge}},
  \bibinfo {author} {\bibfnamefont {H.~D.}\ \bibnamefont {Gliddon}}, \bibinfo
  {author} {\bibfnamefont {D.}~\bibnamefont {Huang}}, \bibinfo {author}
  {\bibfnamefont {G.}~\bibnamefont {Dold}}, \bibinfo {author} {\bibfnamefont
  {E.~R.}\ \bibnamefont {Gray}}, \bibinfo {author} {\bibfnamefont
  {J.}~\bibnamefont {Heaney}}, \bibinfo {author} {\bibfnamefont {P.~J.}\
  \bibnamefont {Dobson}}, \bibinfo {author} {\bibfnamefont {E.}~\bibnamefont
  {Nastouli}}, \bibinfo {author} {\bibfnamefont {J.~J.}\ \bibnamefont
  {Morton}},  \emph {et~al.},\ }\bibfield  {title} {\enquote {\bibinfo {title}
  {Spin-enhanced nanodiamond biosensing for ultrasensitive diagnostics},}\
  }\href@noop {} {\bibfield  {journal} {\bibinfo  {journal} {Nature}\ }\textbf
  {\bibinfo {volume} {587}},\ \bibinfo {pages} {588--593} (\bibinfo {year}
  {2020})}\BibitemShut {NoStop}%
\bibitem [{\citenamefont {Haziza}\ \emph {et~al.}(2017)\citenamefont {Haziza},
  \citenamefont {Mohan}, \citenamefont {Loe-Mie}, \citenamefont
  {Lepagnol-Bestel}, \citenamefont {Massou}, \citenamefont {Adam},
  \citenamefont {Le}, \citenamefont {Viard}, \citenamefont {Plancon},
  \citenamefont {Daudin} \emph {et~al.}}]{haziza2017fluorescent}%
  \BibitemOpen
  \bibfield  {author} {\bibinfo {author} {\bibfnamefont {S.}~\bibnamefont
  {Haziza}}, \bibinfo {author} {\bibfnamefont {N.}~\bibnamefont {Mohan}},
  \bibinfo {author} {\bibfnamefont {Y.}~\bibnamefont {Loe-Mie}}, \bibinfo
  {author} {\bibfnamefont {A.-M.}\ \bibnamefont {Lepagnol-Bestel}}, \bibinfo
  {author} {\bibfnamefont {S.}~\bibnamefont {Massou}}, \bibinfo {author}
  {\bibfnamefont {M.-P.}\ \bibnamefont {Adam}}, \bibinfo {author}
  {\bibfnamefont {X.~L.}\ \bibnamefont {Le}}, \bibinfo {author} {\bibfnamefont
  {J.}~\bibnamefont {Viard}}, \bibinfo {author} {\bibfnamefont
  {C.}~\bibnamefont {Plancon}}, \bibinfo {author} {\bibfnamefont
  {R.}~\bibnamefont {Daudin}},  \emph {et~al.},\ }\bibfield  {title} {\enquote
  {\bibinfo {title} {Fluorescent nanodiamond tracking reveals intraneuronal
  transport abnormalities induced by brain-disease-related genetic risk
  factors},}\ }\href@noop {} {\bibfield  {journal} {\bibinfo  {journal} {Nat.
  Nanotechnol.}\ }\textbf {\bibinfo {volume} {12}},\ \bibinfo {pages}
  {322--328} (\bibinfo {year} {2017})}\BibitemShut {NoStop}%
\bibitem [{\citenamefont {Kucsko}\ \emph {et~al.}(2013)\citenamefont {Kucsko},
  \citenamefont {Maurer}, \citenamefont {Yao}, \citenamefont {Kubo},
  \citenamefont {Noh}, \citenamefont {Lo}, \citenamefont {Park},\ and\
  \citenamefont {Lukin}}]{kucsko2013nanometre}%
  \BibitemOpen
  \bibfield  {author} {\bibinfo {author} {\bibfnamefont {G.}~\bibnamefont
  {Kucsko}}, \bibinfo {author} {\bibfnamefont {P.~C.}\ \bibnamefont {Maurer}},
  \bibinfo {author} {\bibfnamefont {N.~Y.}\ \bibnamefont {Yao}}, \bibinfo
  {author} {\bibfnamefont {M.}~\bibnamefont {Kubo}}, \bibinfo {author}
  {\bibfnamefont {H.~J.}\ \bibnamefont {Noh}}, \bibinfo {author} {\bibfnamefont
  {P.~K.}\ \bibnamefont {Lo}}, \bibinfo {author} {\bibfnamefont
  {H.}~\bibnamefont {Park}}, \ and\ \bibinfo {author} {\bibfnamefont {M.~D.}\
  \bibnamefont {Lukin}},\ }\bibfield  {title} {\enquote {\bibinfo {title}
  {Nanometre-scale thermometry in a living cell},}\ }\href@noop {} {\bibfield
  {journal} {\bibinfo  {journal} {Nature}\ }\textbf {\bibinfo {volume} {500}},\
  \bibinfo {pages} {54--58} (\bibinfo {year} {2013})}\BibitemShut {NoStop}%
\bibitem [{\citenamefont {Simpson}\ \emph {et~al.}(2017)\citenamefont
  {Simpson}, \citenamefont {Morrisroe}, \citenamefont {McCoey}, \citenamefont
  {Lombard}, \citenamefont {Mendis}, \citenamefont {Treussart}, \citenamefont
  {Hall}, \citenamefont {Petrou},\ and\ \citenamefont
  {Hollenberg}}]{simpson2017non}%
  \BibitemOpen
  \bibfield  {author} {\bibinfo {author} {\bibfnamefont {D.~A.}\ \bibnamefont
  {Simpson}}, \bibinfo {author} {\bibfnamefont {E.}~\bibnamefont {Morrisroe}},
  \bibinfo {author} {\bibfnamefont {J.~M.}\ \bibnamefont {McCoey}}, \bibinfo
  {author} {\bibfnamefont {A.~H.}\ \bibnamefont {Lombard}}, \bibinfo {author}
  {\bibfnamefont {D.~C.}\ \bibnamefont {Mendis}}, \bibinfo {author}
  {\bibfnamefont {F.}~\bibnamefont {Treussart}}, \bibinfo {author}
  {\bibfnamefont {L.~T.}\ \bibnamefont {Hall}}, \bibinfo {author}
  {\bibfnamefont {S.}~\bibnamefont {Petrou}}, \ and\ \bibinfo {author}
  {\bibfnamefont {L.~C.}\ \bibnamefont {Hollenberg}},\ }\bibfield  {title}
  {\enquote {\bibinfo {title} {Non-neurotoxic nanodiamond probes for
  intraneuronal temperature mapping},}\ }\href@noop {} {\bibfield  {journal}
  {\bibinfo  {journal} {ACS Nano}\ }\textbf {\bibinfo {volume} {11}},\ \bibinfo
  {pages} {12077--12086} (\bibinfo {year} {2017})}\BibitemShut {NoStop}%
\bibitem [{\citenamefont {Toraille}\ \emph {et~al.}(2018)\citenamefont
  {Toraille}, \citenamefont {A{\"\i}zel}, \citenamefont {Balloul},
  \citenamefont {Vicario}, \citenamefont {Monzel}, \citenamefont {Coppey},
  \citenamefont {Secret}, \citenamefont {Siaugue}, \citenamefont {Sampaio},
  \citenamefont {Rohart} \emph {et~al.}}]{toraille2018optical}%
  \BibitemOpen
  \bibfield  {author} {\bibinfo {author} {\bibfnamefont {L.}~\bibnamefont
  {Toraille}}, \bibinfo {author} {\bibfnamefont {K.}~\bibnamefont
  {A{\"\i}zel}}, \bibinfo {author} {\bibfnamefont {{\'E}.}~\bibnamefont
  {Balloul}}, \bibinfo {author} {\bibfnamefont {C.}~\bibnamefont {Vicario}},
  \bibinfo {author} {\bibfnamefont {C.}~\bibnamefont {Monzel}}, \bibinfo
  {author} {\bibfnamefont {M.}~\bibnamefont {Coppey}}, \bibinfo {author}
  {\bibfnamefont {E.}~\bibnamefont {Secret}}, \bibinfo {author} {\bibfnamefont
  {J.-M.}\ \bibnamefont {Siaugue}}, \bibinfo {author} {\bibfnamefont
  {J.}~\bibnamefont {Sampaio}}, \bibinfo {author} {\bibfnamefont
  {S.}~\bibnamefont {Rohart}},  \emph {et~al.},\ }\bibfield  {title} {\enquote
  {\bibinfo {title} {Optical magnetometry of single biocompatible micromagnets
  for quantitative magnetogenetic and magnetomechanical assays},}\ }\href@noop
  {} {\bibfield  {journal} {\bibinfo  {journal} {Nano Lett.}\ }\textbf
  {\bibinfo {volume} {18}},\ \bibinfo {pages} {7635--7641} (\bibinfo {year}
  {2018})}\BibitemShut {NoStop}%
\bibitem [{\citenamefont {Claveau}, \citenamefont {Bertrand},\ and\
  \citenamefont {Treussart}(2018)}]{claveau2018fluorescent}%
  \BibitemOpen
  \bibfield  {author} {\bibinfo {author} {\bibfnamefont {S.}~\bibnamefont
  {Claveau}}, \bibinfo {author} {\bibfnamefont {J.-R.}\ \bibnamefont
  {Bertrand}}, \ and\ \bibinfo {author} {\bibfnamefont {F.}~\bibnamefont
  {Treussart}},\ }\bibfield  {title} {\enquote {\bibinfo {title} {Fluorescent
  nanodiamond applications for cellular process sensing and cell tracking},}\
  }\href@noop {} {\bibfield  {journal} {\bibinfo  {journal} {Micromachines}\
  }\textbf {\bibinfo {volume} {9}},\ \bibinfo {pages} {247} (\bibinfo {year}
  {2018})}\BibitemShut {NoStop}%
\bibitem [{\citenamefont {Nie}\ \emph {et~al.}(2021{\natexlab{a}})\citenamefont
  {Nie}, \citenamefont {Nusantara}, \citenamefont {Damle}, \citenamefont
  {Sharmin}, \citenamefont {Evans}, \citenamefont {Hemelaar}, \citenamefont
  {van~der Laan}, \citenamefont {Li}, \citenamefont {Martinez}, \citenamefont
  {Vedelaar}, \citenamefont {Chipaux},\ and\ \citenamefont
  {Schirhagl}}]{nie2021quantum}%
  \BibitemOpen
  \bibfield  {author} {\bibinfo {author} {\bibfnamefont {L.}~\bibnamefont
  {Nie}}, \bibinfo {author} {\bibfnamefont {A.}~\bibnamefont {Nusantara}},
  \bibinfo {author} {\bibfnamefont {V.}~\bibnamefont {Damle}}, \bibinfo
  {author} {\bibfnamefont {R.}~\bibnamefont {Sharmin}}, \bibinfo {author}
  {\bibfnamefont {E.}~\bibnamefont {Evans}}, \bibinfo {author} {\bibfnamefont
  {S.}~\bibnamefont {Hemelaar}}, \bibinfo {author} {\bibfnamefont
  {K.}~\bibnamefont {van~der Laan}}, \bibinfo {author} {\bibfnamefont
  {R.}~\bibnamefont {Li}}, \bibinfo {author} {\bibfnamefont {F.~P.}\
  \bibnamefont {Martinez}}, \bibinfo {author} {\bibfnamefont {T.}~\bibnamefont
  {Vedelaar}}, \bibinfo {author} {\bibfnamefont {M.}~\bibnamefont {Chipaux}}, \
  and\ \bibinfo {author} {\bibfnamefont {M.}~\bibnamefont {Schirhagl}},\
  }\bibfield  {title} {\enquote {\bibinfo {title} {Quantum monitoring of
  cellular metabolic activities in single mitochondria},}\ }\href@noop {}
  {\bibfield  {journal} {\bibinfo  {journal} {Sci. Adv.}\ }\textbf {\bibinfo
  {volume} {7}},\ \bibinfo {pages} {eabf0573} (\bibinfo {year}
  {2021}{\natexlab{a}})}\BibitemShut {NoStop}%
\bibitem [{\citenamefont {Davis}\ \emph {et~al.}(2018)\citenamefont {Davis},
  \citenamefont {Ramesh}, \citenamefont {Bhatnagar}, \citenamefont
  {Lee-Gosselin}, \citenamefont {Barry}, \citenamefont {Glenn}, \citenamefont
  {Walsworth},\ and\ \citenamefont {Shapiro}}]{davis2018mapping}%
  \BibitemOpen
  \bibfield  {author} {\bibinfo {author} {\bibfnamefont {H.~C.}\ \bibnamefont
  {Davis}}, \bibinfo {author} {\bibfnamefont {P.}~\bibnamefont {Ramesh}},
  \bibinfo {author} {\bibfnamefont {A.}~\bibnamefont {Bhatnagar}}, \bibinfo
  {author} {\bibfnamefont {A.}~\bibnamefont {Lee-Gosselin}}, \bibinfo {author}
  {\bibfnamefont {J.~F.}\ \bibnamefont {Barry}}, \bibinfo {author}
  {\bibfnamefont {D.~R.}\ \bibnamefont {Glenn}}, \bibinfo {author}
  {\bibfnamefont {R.~L.}\ \bibnamefont {Walsworth}}, \ and\ \bibinfo {author}
  {\bibfnamefont {M.~G.}\ \bibnamefont {Shapiro}},\ }\bibfield  {title}
  {\enquote {\bibinfo {title} {Mapping the microscale origins of magnetic
  resonance image contrast with subcellular diamond magnetometry},}\
  }\href@noop {} {\bibfield  {journal} {\bibinfo  {journal} {Nat. Commun.}\
  }\textbf {\bibinfo {volume} {9}},\ \bibinfo {pages} {1--9} (\bibinfo {year}
  {2018})}\BibitemShut {NoStop}%
\bibitem [{\citenamefont {van~der Laan}\ \emph {et~al.}(2020)\citenamefont
  {van~der Laan}, \citenamefont {Morita}, \citenamefont {Perona-Martinez},\
  and\ \citenamefont {Schirhagl}}]{van2020evaluation}%
  \BibitemOpen
  \bibfield  {author} {\bibinfo {author} {\bibfnamefont {K.~J.}\ \bibnamefont
  {van~der Laan}}, \bibinfo {author} {\bibfnamefont {A.}~\bibnamefont
  {Morita}}, \bibinfo {author} {\bibfnamefont {F.~P.}\ \bibnamefont
  {Perona-Martinez}}, \ and\ \bibinfo {author} {\bibfnamefont {R.}~\bibnamefont
  {Schirhagl}},\ }\bibfield  {title} {\enquote {\bibinfo {title} {Evaluation of
  the oxidative stress response of aging yeast cells in response to
  internalization of fluorescent nanodiamond biosensors},}\ }\href@noop {}
  {\bibfield  {journal} {\bibinfo  {journal} {Nanomaterials}\ }\textbf
  {\bibinfo {volume} {10}},\ \bibinfo {pages} {372} (\bibinfo {year}
  {2020})}\BibitemShut {NoStop}%
\bibitem [{\citenamefont {Fujiwara}\ \emph {et~al.}(2020)\citenamefont
  {Fujiwara}, \citenamefont {Sun}, \citenamefont {Dohms}, \citenamefont
  {Nishimura}, \citenamefont {Suto}, \citenamefont {Takezawa}, \citenamefont
  {Oshimi}, \citenamefont {Zhao}, \citenamefont {Sadzak}, \citenamefont
  {Umehara} \emph {et~al.}}]{fujiwara2020real}%
  \BibitemOpen
  \bibfield  {author} {\bibinfo {author} {\bibfnamefont {M.}~\bibnamefont
  {Fujiwara}}, \bibinfo {author} {\bibfnamefont {S.}~\bibnamefont {Sun}},
  \bibinfo {author} {\bibfnamefont {A.}~\bibnamefont {Dohms}}, \bibinfo
  {author} {\bibfnamefont {Y.}~\bibnamefont {Nishimura}}, \bibinfo {author}
  {\bibfnamefont {K.}~\bibnamefont {Suto}}, \bibinfo {author} {\bibfnamefont
  {Y.}~\bibnamefont {Takezawa}}, \bibinfo {author} {\bibfnamefont
  {K.}~\bibnamefont {Oshimi}}, \bibinfo {author} {\bibfnamefont
  {L.}~\bibnamefont {Zhao}}, \bibinfo {author} {\bibfnamefont {N.}~\bibnamefont
  {Sadzak}}, \bibinfo {author} {\bibfnamefont {Y.}~\bibnamefont {Umehara}},
  \emph {et~al.},\ }\bibfield  {title} {\enquote {\bibinfo {title} {Real-time
  nanodiamond thermometry probing in vivo thermogenic responses},}\ }\href@noop
  {} {\bibfield  {journal} {\bibinfo  {journal} {Sci. Adv.}\ }\textbf {\bibinfo
  {volume} {6}},\ \bibinfo {pages} {eaba9636} (\bibinfo {year}
  {2020})}\BibitemShut {NoStop}%
\bibitem [{\citenamefont {Rondin}\ \emph {et~al.}(2014)\citenamefont {Rondin},
  \citenamefont {Tetienne}, \citenamefont {Hingant}, \citenamefont {Roch},
  \citenamefont {Maletinsky},\ and\ \citenamefont
  {Jacques}}]{rondin2014magnetometry}%
  \BibitemOpen
  \bibfield  {author} {\bibinfo {author} {\bibfnamefont {L.}~\bibnamefont
  {Rondin}}, \bibinfo {author} {\bibfnamefont {J.-P.}\ \bibnamefont
  {Tetienne}}, \bibinfo {author} {\bibfnamefont {T.}~\bibnamefont {Hingant}},
  \bibinfo {author} {\bibfnamefont {J.-F.}\ \bibnamefont {Roch}}, \bibinfo
  {author} {\bibfnamefont {P.}~\bibnamefont {Maletinsky}}, \ and\ \bibinfo
  {author} {\bibfnamefont {V.}~\bibnamefont {Jacques}},\ }\bibfield  {title}
  {\enquote {\bibinfo {title} {Magnetometry with nitrogen-vacancy defects in
  diamond},}\ }\href@noop {} {\bibfield  {journal} {\bibinfo  {journal} {Rep.
  Prog. Phys.}\ }\textbf {\bibinfo {volume} {77}},\ \bibinfo {pages} {056503}
  (\bibinfo {year} {2014})}\BibitemShut {NoStop}%
\bibitem [{\citenamefont {Maclaurin}\ \emph {et~al.}(2013)\citenamefont
  {Maclaurin}, \citenamefont {Hall}, \citenamefont {Martin},\ and\
  \citenamefont {Hollenberg}}]{maclaurin2013nanoscale}%
  \BibitemOpen
  \bibfield  {author} {\bibinfo {author} {\bibfnamefont {D.}~\bibnamefont
  {Maclaurin}}, \bibinfo {author} {\bibfnamefont {L.}~\bibnamefont {Hall}},
  \bibinfo {author} {\bibfnamefont {A.}~\bibnamefont {Martin}}, \ and\ \bibinfo
  {author} {\bibfnamefont {L.}~\bibnamefont {Hollenberg}},\ }\bibfield  {title}
  {\enquote {\bibinfo {title} {Nanoscale magnetometry through quantum control
  of nitrogen--vacancy centres in rotationally diffusing nanodiamonds},}\
  }\href@noop {} {\bibfield  {journal} {\bibinfo  {journal} {New J. Phys.}\
  }\textbf {\bibinfo {volume} {15}},\ \bibinfo {pages} {013041} (\bibinfo
  {year} {2013})}\BibitemShut {NoStop}%
\bibitem [{\citenamefont {Horowitz}\ \emph {et~al.}(2012)\citenamefont
  {Horowitz}, \citenamefont {Alem{\'a}n}, \citenamefont {Christle},
  \citenamefont {Cleland},\ and\ \citenamefont
  {Awschalom}}]{horowitz2012electron}%
  \BibitemOpen
  \bibfield  {author} {\bibinfo {author} {\bibfnamefont {V.~R.}\ \bibnamefont
  {Horowitz}}, \bibinfo {author} {\bibfnamefont {B.~J.}\ \bibnamefont
  {Alem{\'a}n}}, \bibinfo {author} {\bibfnamefont {D.~J.}\ \bibnamefont
  {Christle}}, \bibinfo {author} {\bibfnamefont {A.~N.}\ \bibnamefont
  {Cleland}}, \ and\ \bibinfo {author} {\bibfnamefont {D.~D.}\ \bibnamefont
  {Awschalom}},\ }\bibfield  {title} {\enquote {\bibinfo {title} {Electron spin
  resonance of nitrogen-vacancy centers in optically trapped nanodiamonds},}\
  }\href@noop {} {\bibfield  {journal} {\bibinfo  {journal} {Proc. Natl. Acad.
  Sci. U.S.A.}\ }\textbf {\bibinfo {volume} {109}},\ \bibinfo {pages}
  {13493--13497} (\bibinfo {year} {2012})}\BibitemShut {NoStop}%
\bibitem [{\citenamefont {Dolde}\ \emph {et~al.}(2011)\citenamefont {Dolde},
  \citenamefont {Fedder}, \citenamefont {Doherty}, \citenamefont {N{\"o}bauer},
  \citenamefont {Rempp}, \citenamefont {Balasubramanian}, \citenamefont {Wolf},
  \citenamefont {Reinhard}, \citenamefont {Hollenberg}, \citenamefont {Jelezko}
  \emph {et~al.}}]{dolde2011electric}%
  \BibitemOpen
  \bibfield  {author} {\bibinfo {author} {\bibfnamefont {F.}~\bibnamefont
  {Dolde}}, \bibinfo {author} {\bibfnamefont {H.}~\bibnamefont {Fedder}},
  \bibinfo {author} {\bibfnamefont {M.~W.}\ \bibnamefont {Doherty}}, \bibinfo
  {author} {\bibfnamefont {T.}~\bibnamefont {N{\"o}bauer}}, \bibinfo {author}
  {\bibfnamefont {F.}~\bibnamefont {Rempp}}, \bibinfo {author} {\bibfnamefont
  {G.}~\bibnamefont {Balasubramanian}}, \bibinfo {author} {\bibfnamefont
  {T.}~\bibnamefont {Wolf}}, \bibinfo {author} {\bibfnamefont {F.}~\bibnamefont
  {Reinhard}}, \bibinfo {author} {\bibfnamefont {L.~C.}\ \bibnamefont
  {Hollenberg}}, \bibinfo {author} {\bibfnamefont {F.}~\bibnamefont {Jelezko}},
   \emph {et~al.},\ }\bibfield  {title} {\enquote {\bibinfo {title}
  {Electric-field sensing using single diamond spins},}\ }\href@noop {}
  {\bibfield  {journal} {\bibinfo  {journal} {Nat. Phys.}\ }\textbf {\bibinfo
  {volume} {7}},\ \bibinfo {pages} {459--463} (\bibinfo {year}
  {2011})}\BibitemShut {NoStop}%
\bibitem [{\citenamefont {Iwasaki}\ \emph {et~al.}(2017)\citenamefont
  {Iwasaki}, \citenamefont {Naruki}, \citenamefont {Tahara}, \citenamefont
  {Makino}, \citenamefont {Kato}, \citenamefont {Ogura}, \citenamefont
  {Takeuchi}, \citenamefont {Yamasaki},\ and\ \citenamefont
  {Hatano}}]{iwasaki2017direct}%
  \BibitemOpen
  \bibfield  {author} {\bibinfo {author} {\bibfnamefont {T.}~\bibnamefont
  {Iwasaki}}, \bibinfo {author} {\bibfnamefont {W.}~\bibnamefont {Naruki}},
  \bibinfo {author} {\bibfnamefont {K.}~\bibnamefont {Tahara}}, \bibinfo
  {author} {\bibfnamefont {T.}~\bibnamefont {Makino}}, \bibinfo {author}
  {\bibfnamefont {H.}~\bibnamefont {Kato}}, \bibinfo {author} {\bibfnamefont
  {M.}~\bibnamefont {Ogura}}, \bibinfo {author} {\bibfnamefont
  {D.}~\bibnamefont {Takeuchi}}, \bibinfo {author} {\bibfnamefont
  {S.}~\bibnamefont {Yamasaki}}, \ and\ \bibinfo {author} {\bibfnamefont
  {M.}~\bibnamefont {Hatano}},\ }\bibfield  {title} {\enquote {\bibinfo {title}
  {Direct nanoscale sensing of the internal electric field in operating
  semiconductor devices using single electron spins},}\ }\href@noop {}
  {\bibfield  {journal} {\bibinfo  {journal} {ACS Nano}\ }\textbf {\bibinfo
  {volume} {11}},\ \bibinfo {pages} {1238--1245} (\bibinfo {year}
  {2017})}\BibitemShut {NoStop}%
\bibitem [{\citenamefont {Bian}\ \emph {et~al.}(2021)\citenamefont {Bian},
  \citenamefont {Zheng}, \citenamefont {Zeng}, \citenamefont {Chen},
  \citenamefont {St{\"o}hr}, \citenamefont {Denisenko}, \citenamefont {Yang},
  \citenamefont {Wrachtrup},\ and\ \citenamefont {Jiang}}]{bian2021nanoscale}%
  \BibitemOpen
  \bibfield  {author} {\bibinfo {author} {\bibfnamefont {K.}~\bibnamefont
  {Bian}}, \bibinfo {author} {\bibfnamefont {W.}~\bibnamefont {Zheng}},
  \bibinfo {author} {\bibfnamefont {X.}~\bibnamefont {Zeng}}, \bibinfo {author}
  {\bibfnamefont {X.}~\bibnamefont {Chen}}, \bibinfo {author} {\bibfnamefont
  {R.}~\bibnamefont {St{\"o}hr}}, \bibinfo {author} {\bibfnamefont
  {A.}~\bibnamefont {Denisenko}}, \bibinfo {author} {\bibfnamefont
  {S.}~\bibnamefont {Yang}}, \bibinfo {author} {\bibfnamefont {J.}~\bibnamefont
  {Wrachtrup}}, \ and\ \bibinfo {author} {\bibfnamefont {Y.}~\bibnamefont
  {Jiang}},\ }\bibfield  {title} {\enquote {\bibinfo {title} {Nanoscale
  electric-field imaging based on a quantum sensor and its charge-state control
  under ambient condition},}\ }\href@noop {} {\bibfield  {journal} {\bibinfo
  {journal} {Nat. Commun.}\ }\textbf {\bibinfo {volume} {12}},\ \bibinfo
  {pages} {2457} (\bibinfo {year} {2021})}\BibitemShut {NoStop}%
\bibitem [{\citenamefont {Neumann}\ \emph {et~al.}(2013)\citenamefont
  {Neumann}, \citenamefont {Jakobi}, \citenamefont {Dolde}, \citenamefont
  {Burk}, \citenamefont {Reuter}, \citenamefont {Waldherr}, \citenamefont
  {Honert}, \citenamefont {Wolf}, \citenamefont {Brunner}, \citenamefont {Shim}
  \emph {et~al.}}]{neumann2013high}%
  \BibitemOpen
  \bibfield  {author} {\bibinfo {author} {\bibfnamefont {P.}~\bibnamefont
  {Neumann}}, \bibinfo {author} {\bibfnamefont {I.}~\bibnamefont {Jakobi}},
  \bibinfo {author} {\bibfnamefont {F.}~\bibnamefont {Dolde}}, \bibinfo
  {author} {\bibfnamefont {C.}~\bibnamefont {Burk}}, \bibinfo {author}
  {\bibfnamefont {R.}~\bibnamefont {Reuter}}, \bibinfo {author} {\bibfnamefont
  {G.}~\bibnamefont {Waldherr}}, \bibinfo {author} {\bibfnamefont
  {J.}~\bibnamefont {Honert}}, \bibinfo {author} {\bibfnamefont
  {T.}~\bibnamefont {Wolf}}, \bibinfo {author} {\bibfnamefont {A.}~\bibnamefont
  {Brunner}}, \bibinfo {author} {\bibfnamefont {J.~H.}\ \bibnamefont {Shim}},
  \emph {et~al.},\ }\bibfield  {title} {\enquote {\bibinfo {title}
  {High-precision nanoscale temperature sensing using single defects in
  diamond},}\ }\href@noop {} {\bibfield  {journal} {\bibinfo  {journal} {Nano
  Lett.}\ }\textbf {\bibinfo {volume} {13}},\ \bibinfo {pages} {2738--2742}
  (\bibinfo {year} {2013})}\BibitemShut {NoStop}%
\bibitem [{\citenamefont {Fujiwara}\ and\ \citenamefont
  {Shikano}(2021)}]{fujiwara2021diamond}%
  \BibitemOpen
  \bibfield  {author} {\bibinfo {author} {\bibfnamefont {M.}~\bibnamefont
  {Fujiwara}}\ and\ \bibinfo {author} {\bibfnamefont {Y.}~\bibnamefont
  {Shikano}},\ }\bibfield  {title} {\enquote {\bibinfo {title} {Diamond quantum
  thermometry: From foundations to applications},}\ }\href@noop {} {\bibfield
  {journal} {\bibinfo  {journal} {Nanotechnology}\ }\textbf {\bibinfo {volume}
  {32}},\ \bibinfo {pages} {482002} (\bibinfo {year} {2021})}\BibitemShut
  {NoStop}%
\bibitem [{\citenamefont {Wang}\ \emph {et~al.}(2018)\citenamefont {Wang},
  \citenamefont {Liu}, \citenamefont {Leong}, \citenamefont {Zeng},
  \citenamefont {Feng}, \citenamefont {Li}, \citenamefont {Dolde},
  \citenamefont {Fedder}, \citenamefont {Wrachtrup}, \citenamefont {Cui} \emph
  {et~al.}}]{wang2018magnetic}%
  \BibitemOpen
  \bibfield  {author} {\bibinfo {author} {\bibfnamefont {N.}~\bibnamefont
  {Wang}}, \bibinfo {author} {\bibfnamefont {G.-Q.}\ \bibnamefont {Liu}},
  \bibinfo {author} {\bibfnamefont {W.-H.}\ \bibnamefont {Leong}}, \bibinfo
  {author} {\bibfnamefont {H.}~\bibnamefont {Zeng}}, \bibinfo {author}
  {\bibfnamefont {X.}~\bibnamefont {Feng}}, \bibinfo {author} {\bibfnamefont
  {S.-H.}\ \bibnamefont {Li}}, \bibinfo {author} {\bibfnamefont
  {F.}~\bibnamefont {Dolde}}, \bibinfo {author} {\bibfnamefont
  {H.}~\bibnamefont {Fedder}}, \bibinfo {author} {\bibfnamefont
  {J.}~\bibnamefont {Wrachtrup}}, \bibinfo {author} {\bibfnamefont {X.-D.}\
  \bibnamefont {Cui}},  \emph {et~al.},\ }\bibfield  {title} {\enquote
  {\bibinfo {title} {Magnetic criticality enhanced hybrid nanodiamond
  thermometer under ambient conditions},}\ }\href@noop {} {\bibfield  {journal}
  {\bibinfo  {journal} {Phys. Rev. X.}\ }\textbf {\bibinfo {volume} {8}},\
  \bibinfo {pages} {011042} (\bibinfo {year} {2018})}\BibitemShut {NoStop}%
\bibitem [{\citenamefont {Fujisaku}\ \emph {et~al.}(2019)\citenamefont
  {Fujisaku}, \citenamefont {Tanabe}, \citenamefont {Onoda}, \citenamefont
  {Kubota}, \citenamefont {Segawa}, \citenamefont {So}, \citenamefont
  {Ohshima}, \citenamefont {Hamachi}, \citenamefont {Shirakawa},\ and\
  \citenamefont {Igarashi}}]{fujisaku2019ph}%
  \BibitemOpen
  \bibfield  {author} {\bibinfo {author} {\bibfnamefont {T.}~\bibnamefont
  {Fujisaku}}, \bibinfo {author} {\bibfnamefont {R.}~\bibnamefont {Tanabe}},
  \bibinfo {author} {\bibfnamefont {S.}~\bibnamefont {Onoda}}, \bibinfo
  {author} {\bibfnamefont {R.}~\bibnamefont {Kubota}}, \bibinfo {author}
  {\bibfnamefont {T.~F.}\ \bibnamefont {Segawa}}, \bibinfo {author}
  {\bibfnamefont {F.~T.-K.}\ \bibnamefont {So}}, \bibinfo {author}
  {\bibfnamefont {T.}~\bibnamefont {Ohshima}}, \bibinfo {author} {\bibfnamefont
  {I.}~\bibnamefont {Hamachi}}, \bibinfo {author} {\bibfnamefont
  {M.}~\bibnamefont {Shirakawa}}, \ and\ \bibinfo {author} {\bibfnamefont
  {R.}~\bibnamefont {Igarashi}},\ }\bibfield  {title} {\enquote {\bibinfo
  {title} {ph nanosensor using electronic spins in diamond},}\ }\href@noop {}
  {\bibfield  {journal} {\bibinfo  {journal} {ACS nano}\ }\textbf {\bibinfo
  {volume} {13}},\ \bibinfo {pages} {11726--11732} (\bibinfo {year}
  {2019})}\BibitemShut {NoStop}%
\bibitem [{\citenamefont {Nie}\ \emph {et~al.}(2022)\citenamefont {Nie},
  \citenamefont {Nusantara}, \citenamefont {Damle}, \citenamefont {Baranov},
  \citenamefont {Chipaux}, \citenamefont {Reyes-San-Martin}, \citenamefont
  {Hamoh}, \citenamefont {Epperla}, \citenamefont {Guricova}, \citenamefont
  {Cigler}, \citenamefont {van~den Bogaart},\ and\ \citenamefont
  {Schirhagl}}]{Nie-NanoLett-2022}%
  \BibitemOpen
  \bibfield  {author} {\bibinfo {author} {\bibfnamefont {L.}~\bibnamefont
  {Nie}}, \bibinfo {author} {\bibfnamefont {A.~C.}\ \bibnamefont {Nusantara}},
  \bibinfo {author} {\bibfnamefont {V.~G.}\ \bibnamefont {Damle}}, \bibinfo
  {author} {\bibfnamefont {M.~V.}\ \bibnamefont {Baranov}}, \bibinfo {author}
  {\bibfnamefont {M.}~\bibnamefont {Chipaux}}, \bibinfo {author} {\bibfnamefont
  {C.}~\bibnamefont {Reyes-San-Martin}}, \bibinfo {author} {\bibfnamefont
  {T.}~\bibnamefont {Hamoh}}, \bibinfo {author} {\bibfnamefont {C.~P.}\
  \bibnamefont {Epperla}}, \bibinfo {author} {\bibfnamefont {M.}~\bibnamefont
  {Guricova}}, \bibinfo {author} {\bibfnamefont {P.}~\bibnamefont {Cigler}},
  \bibinfo {author} {\bibfnamefont {G.}~\bibnamefont {van~den Bogaart}}, \ and\
  \bibinfo {author} {\bibfnamefont {R.}~\bibnamefont {Schirhagl}},\ }\bibfield
  {title} {\enquote {\bibinfo {title} {Quantum sensing of free radicals in
  primary human dendritic cells},}\ }\href {\doibase
  10.1021/acs.nanolett.1c03021} {\bibfield  {journal} {\bibinfo  {journal}
  {Nano Letters}\ }\textbf {\bibinfo {volume} {22}},\ \bibinfo {pages}
  {1818--1825} (\bibinfo {year} {2022})},\ \bibinfo {note} {pMID: 34929080},\
  \Eprint {http://arxiv.org/abs/https://doi.org/10.1021/acs.nanolett.1c03021}
  {https://doi.org/10.1021/acs.nanolett.1c03021} \BibitemShut {NoStop}%
\bibitem [{\citenamefont {Allert}\ \emph {et~al.}(2022)\citenamefont {Allert},
  \citenamefont {Bruckmaier}, \citenamefont {Neuling}, \citenamefont
  {Freire-Moschovitis}, \citenamefont {Liu}, \citenamefont {Schrepel},
  \citenamefont {Sch{\"a}tzle}, \citenamefont {Knittel}, \citenamefont
  {Hermans},\ and\ \citenamefont {Bucher}}]{allert2022microfluidic}%
  \BibitemOpen
  \bibfield  {author} {\bibinfo {author} {\bibfnamefont {R.~D.}\ \bibnamefont
  {Allert}}, \bibinfo {author} {\bibfnamefont {F.}~\bibnamefont {Bruckmaier}},
  \bibinfo {author} {\bibfnamefont {N.~R.}\ \bibnamefont {Neuling}}, \bibinfo
  {author} {\bibfnamefont {F.~A.}\ \bibnamefont {Freire-Moschovitis}}, \bibinfo
  {author} {\bibfnamefont {K.~S.}\ \bibnamefont {Liu}}, \bibinfo {author}
  {\bibfnamefont {C.}~\bibnamefont {Schrepel}}, \bibinfo {author}
  {\bibfnamefont {P.}~\bibnamefont {Sch{\"a}tzle}}, \bibinfo {author}
  {\bibfnamefont {P.}~\bibnamefont {Knittel}}, \bibinfo {author} {\bibfnamefont
  {M.}~\bibnamefont {Hermans}}, \ and\ \bibinfo {author} {\bibfnamefont
  {D.~B.}\ \bibnamefont {Bucher}},\ }\bibfield  {title} {\enquote {\bibinfo
  {title} {Microfluidic quantum sensing platform for lab-on-a-chip
  applications},}\ }\href@noop {} {\bibfield  {journal} {\bibinfo  {journal}
  {Lab on a Chip}\ }\textbf {\bibinfo {volume} {22}},\ \bibinfo {pages}
  {4831--4840} (\bibinfo {year} {2022})}\BibitemShut {NoStop}%
\bibitem [{\citenamefont {Oshimi}\ \emph {et~al.}(2022)\citenamefont {Oshimi},
  \citenamefont {Nishimura}, \citenamefont {Matsubara}, \citenamefont {Tanaka},
  \citenamefont {Shikoh}, \citenamefont {Zhao}, \citenamefont {Zou},
  \citenamefont {Komatsu}, \citenamefont {Ikado}, \citenamefont {Takezawa}
  \emph {et~al.}}]{oshimi2022glass}%
  \BibitemOpen
  \bibfield  {author} {\bibinfo {author} {\bibfnamefont {K.}~\bibnamefont
  {Oshimi}}, \bibinfo {author} {\bibfnamefont {Y.}~\bibnamefont {Nishimura}},
  \bibinfo {author} {\bibfnamefont {T.}~\bibnamefont {Matsubara}}, \bibinfo
  {author} {\bibfnamefont {M.}~\bibnamefont {Tanaka}}, \bibinfo {author}
  {\bibfnamefont {E.}~\bibnamefont {Shikoh}}, \bibinfo {author} {\bibfnamefont
  {L.}~\bibnamefont {Zhao}}, \bibinfo {author} {\bibfnamefont {Y.}~\bibnamefont
  {Zou}}, \bibinfo {author} {\bibfnamefont {N.}~\bibnamefont {Komatsu}},
  \bibinfo {author} {\bibfnamefont {Y.}~\bibnamefont {Ikado}}, \bibinfo
  {author} {\bibfnamefont {Y.}~\bibnamefont {Takezawa}},  \emph {et~al.},\
  }\bibfield  {title} {\enquote {\bibinfo {title} {Glass-patternable
  notch-shaped microwave architecture for on-chip spin detection in biological
  samples},}\ }\href@noop {} {\bibfield  {journal} {\bibinfo  {journal} {Lab on
  a Chip}\ }\textbf {\bibinfo {volume} {22}},\ \bibinfo {pages} {2519--2530}
  (\bibinfo {year} {2022})}\BibitemShut {NoStop}%
\bibitem [{\citenamefont {Le~Sage}\ \emph {et~al.}(2012)\citenamefont
  {Le~Sage}, \citenamefont {Pham}, \citenamefont {Bar-Gill}, \citenamefont
  {Belthangady}, \citenamefont {Lukin}, \citenamefont {Yacoby},\ and\
  \citenamefont {Walsworth}}]{PhysRevB.85.121202}%
  \BibitemOpen
  \bibfield  {author} {\bibinfo {author} {\bibfnamefont {D.}~\bibnamefont
  {Le~Sage}}, \bibinfo {author} {\bibfnamefont {L.~M.}\ \bibnamefont {Pham}},
  \bibinfo {author} {\bibfnamefont {N.}~\bibnamefont {Bar-Gill}}, \bibinfo
  {author} {\bibfnamefont {C.}~\bibnamefont {Belthangady}}, \bibinfo {author}
  {\bibfnamefont {M.~D.}\ \bibnamefont {Lukin}}, \bibinfo {author}
  {\bibfnamefont {A.}~\bibnamefont {Yacoby}}, \ and\ \bibinfo {author}
  {\bibfnamefont {R.~L.}\ \bibnamefont {Walsworth}},\ }\bibfield  {title}
  {\enquote {\bibinfo {title} {Efficient photon detection from color centers in
  a diamond optical waveguide},}\ }\href {\doibase 10.1103/PhysRevB.85.121202}
  {\bibfield  {journal} {\bibinfo  {journal} {Phys. Rev. B}\ }\textbf {\bibinfo
  {volume} {85}},\ \bibinfo {pages} {121202} (\bibinfo {year}
  {2012})}\BibitemShut {NoStop}%
\bibitem [{\citenamefont {Losero}\ \emph {et~al.}(2023)\citenamefont {Losero},
  \citenamefont {Jagannath}, \citenamefont {Pezzoli}, \citenamefont {Goblot},
  \citenamefont {Babashah}, \citenamefont {Lashuel}, \citenamefont {Galland},\
  and\ \citenamefont {Quack}}]{losero2023neuronal}%
  \BibitemOpen
  \bibfield  {author} {\bibinfo {author} {\bibfnamefont {E.}~\bibnamefont
  {Losero}}, \bibinfo {author} {\bibfnamefont {S.}~\bibnamefont {Jagannath}},
  \bibinfo {author} {\bibfnamefont {M.}~\bibnamefont {Pezzoli}}, \bibinfo
  {author} {\bibfnamefont {V.}~\bibnamefont {Goblot}}, \bibinfo {author}
  {\bibfnamefont {H.}~\bibnamefont {Babashah}}, \bibinfo {author}
  {\bibfnamefont {H.~A.}\ \bibnamefont {Lashuel}}, \bibinfo {author}
  {\bibfnamefont {C.}~\bibnamefont {Galland}}, \ and\ \bibinfo {author}
  {\bibfnamefont {N.}~\bibnamefont {Quack}},\ }\bibfield  {title} {\enquote
  {\bibinfo {title} {Neuronal growth on high-aspect-ratio diamond nanopillar
  arrays for biosensing applications},}\ }\href@noop {} {\bibfield  {journal}
  {\bibinfo  {journal} {Scientific Reports}\ }\textbf {\bibinfo {volume}
  {13}},\ \bibinfo {pages} {5909} (\bibinfo {year} {2023})}\BibitemShut
  {NoStop}%
\bibitem [{\citenamefont {Nie}\ \emph {et~al.}(2021{\natexlab{b}})\citenamefont
  {Nie}, \citenamefont {Zhang}, \citenamefont {Li}, \citenamefont {van Rijn},\
  and\ \citenamefont {Schirhagl}}]{nano11071837}%
  \BibitemOpen
  \bibfield  {author} {\bibinfo {author} {\bibfnamefont {L.}~\bibnamefont
  {Nie}}, \bibinfo {author} {\bibfnamefont {Y.}~\bibnamefont {Zhang}}, \bibinfo
  {author} {\bibfnamefont {L.}~\bibnamefont {Li}}, \bibinfo {author}
  {\bibfnamefont {P.}~\bibnamefont {van Rijn}}, \ and\ \bibinfo {author}
  {\bibfnamefont {R.}~\bibnamefont {Schirhagl}},\ }\bibfield  {title} {\enquote
  {\bibinfo {title} {ph sensitive dextran coated fluorescent nanodiamonds as a
  biomarker for hela cells endocytic pathway and increased cellular uptake},}\
  }\href {\doibase 10.3390/nano11071837} {\bibfield  {journal} {\bibinfo
  {journal} {Nanomaterials}\ }\textbf {\bibinfo {volume} {11}} (\bibinfo {year}
  {2021}{\natexlab{b}}),\ 10.3390/nano11071837}\BibitemShut {NoStop}%
\bibitem [{\citenamefont {Andrich}\ \emph {et~al.}(2014)\citenamefont
  {Andrich}, \citenamefont {Alemán}, \citenamefont {Lee}, \citenamefont
  {Ohno}, \citenamefont {de~las Casas}, \citenamefont {Heremans}, \citenamefont
  {Hu},\ and\ \citenamefont {Awschalom}}]{Andrich-NanoLett-2014}%
  \BibitemOpen
  \bibfield  {author} {\bibinfo {author} {\bibfnamefont {P.}~\bibnamefont
  {Andrich}}, \bibinfo {author} {\bibfnamefont {B.~J.}\ \bibnamefont
  {Alemán}}, \bibinfo {author} {\bibfnamefont {J.~C.}\ \bibnamefont {Lee}},
  \bibinfo {author} {\bibfnamefont {K.}~\bibnamefont {Ohno}}, \bibinfo {author}
  {\bibfnamefont {C.~F.}\ \bibnamefont {de~las Casas}}, \bibinfo {author}
  {\bibfnamefont {F.~J.}\ \bibnamefont {Heremans}}, \bibinfo {author}
  {\bibfnamefont {E.~L.}\ \bibnamefont {Hu}}, \ and\ \bibinfo {author}
  {\bibfnamefont {D.~D.}\ \bibnamefont {Awschalom}},\ }\bibfield  {title}
  {\enquote {\bibinfo {title} {Engineered micro- and nanoscale diamonds as
  mobile probes for high-resolution sensing in fluid},}\ }\href {\doibase
  10.1021/nl501208s} {\bibfield  {journal} {\bibinfo  {journal} {Nano Letters}\
  }\textbf {\bibinfo {volume} {14}},\ \bibinfo {pages} {4959--4964} (\bibinfo
  {year} {2014})},\ \bibinfo {note} {pMID: 25076417},\ \Eprint
  {http://arxiv.org/abs/https://doi.org/10.1021/nl501208s}
  {https://doi.org/10.1021/nl501208s} \BibitemShut {NoStop}%
\bibitem [{\citenamefont {Lim}\ \emph {et~al.}(2015)\citenamefont {Lim},
  \citenamefont {Ropp}, \citenamefont {Shapiro}, \citenamefont {Taylor},\ and\
  \citenamefont {Waks}}]{Lim-NanoLett-2015}%
  \BibitemOpen
  \bibfield  {author} {\bibinfo {author} {\bibfnamefont {K.}~\bibnamefont
  {Lim}}, \bibinfo {author} {\bibfnamefont {C.}~\bibnamefont {Ropp}}, \bibinfo
  {author} {\bibfnamefont {B.}~\bibnamefont {Shapiro}}, \bibinfo {author}
  {\bibfnamefont {J.~M.}\ \bibnamefont {Taylor}}, \ and\ \bibinfo {author}
  {\bibfnamefont {E.}~\bibnamefont {Waks}},\ }\bibfield  {title} {\enquote
  {\bibinfo {title} {Scanning localized magnetic fields in a microfluidic
  device with a single nitrogen vacancy center},}\ }\href {\doibase
  10.1021/nl503280u} {\bibfield  {journal} {\bibinfo  {journal} {Nano Letters}\
  }\textbf {\bibinfo {volume} {15}},\ \bibinfo {pages} {1481--1486} (\bibinfo
  {year} {2015})},\ \bibinfo {note} {pMID: 25654268},\ \Eprint
  {http://arxiv.org/abs/https://doi.org/10.1021/nl503280u}
  {https://doi.org/10.1021/nl503280u} \BibitemShut {NoStop}%
\bibitem [{\citenamefont {Smits}\ \emph {et~al.}(2019)\citenamefont {Smits},
  \citenamefont {Damron}, \citenamefont {Kehayias}, \citenamefont {McDowell},
  \citenamefont {Mosavian}, \citenamefont {Fescenko}, \citenamefont {Ristoff},
  \citenamefont {Laraoui}, \citenamefont {Jarmola},\ and\ \citenamefont
  {Acosta}}]{Smits-SciAdv-2019}%
  \BibitemOpen
  \bibfield  {author} {\bibinfo {author} {\bibfnamefont {J.}~\bibnamefont
  {Smits}}, \bibinfo {author} {\bibfnamefont {J.~T.}\ \bibnamefont {Damron}},
  \bibinfo {author} {\bibfnamefont {P.}~\bibnamefont {Kehayias}}, \bibinfo
  {author} {\bibfnamefont {A.~F.}\ \bibnamefont {McDowell}}, \bibinfo {author}
  {\bibfnamefont {N.}~\bibnamefont {Mosavian}}, \bibinfo {author}
  {\bibfnamefont {I.}~\bibnamefont {Fescenko}}, \bibinfo {author}
  {\bibfnamefont {N.}~\bibnamefont {Ristoff}}, \bibinfo {author} {\bibfnamefont
  {A.}~\bibnamefont {Laraoui}}, \bibinfo {author} {\bibfnamefont
  {A.}~\bibnamefont {Jarmola}}, \ and\ \bibinfo {author} {\bibfnamefont
  {V.~M.}\ \bibnamefont {Acosta}},\ }\bibfield  {title} {\enquote {\bibinfo
  {title} {Two-dimensional nuclear magnetic resonance spectroscopy with a
  microfluidic diamond quantum sensor},}\ }\href {\doibase
  10.1126/sciadv.aaw7895} {\bibfield  {journal} {\bibinfo  {journal} {Science
  Advances}\ }\textbf {\bibinfo {volume} {5}},\ \bibinfo {pages} {eaaw7895}
  (\bibinfo {year} {2019})},\ \Eprint
  {http://arxiv.org/abs/https://www.science.org/doi/pdf/10.1126/sciadv.aaw7895}
  {https://www.science.org/doi/pdf/10.1126/sciadv.aaw7895} \BibitemShut
  {NoStop}%
\bibitem [{\citenamefont {Liu}\ \emph {et~al.}(2022)\citenamefont {Liu},
  \citenamefont {Ma}, \citenamefont {Rizzato}, \citenamefont {Semrau},
  \citenamefont {Henning}, \citenamefont {Sharp}, \citenamefont {Fischer},\
  and\ \citenamefont {Bucher}}]{Liu-MOF-NanoLett-2022}%
  \BibitemOpen
  \bibfield  {author} {\bibinfo {author} {\bibfnamefont {K.~S.}\ \bibnamefont
  {Liu}}, \bibinfo {author} {\bibfnamefont {X.}~\bibnamefont {Ma}}, \bibinfo
  {author} {\bibfnamefont {R.}~\bibnamefont {Rizzato}}, \bibinfo {author}
  {\bibfnamefont {A.~L.}\ \bibnamefont {Semrau}}, \bibinfo {author}
  {\bibfnamefont {A.}~\bibnamefont {Henning}}, \bibinfo {author} {\bibfnamefont
  {I.~D.}\ \bibnamefont {Sharp}}, \bibinfo {author} {\bibfnamefont {R.~A.}\
  \bibnamefont {Fischer}}, \ and\ \bibinfo {author} {\bibfnamefont {D.~B.}\
  \bibnamefont {Bucher}},\ }\bibfield  {title} {\enquote {\bibinfo {title}
  {Using metal–organic frameworks to confine liquid samples for nanoscale
  nv-nmr},}\ }\href {\doibase 10.1021/acs.nanolett.2c03069} {\bibfield
  {journal} {\bibinfo  {journal} {Nano Letters}\ }\textbf {\bibinfo {volume}
  {22}},\ \bibinfo {pages} {9876--9882} (\bibinfo {year} {2022})},\ \bibinfo
  {note} {pMID: 36480706},\ \Eprint
  {http://arxiv.org/abs/https://doi.org/10.1021/acs.nanolett.2c03069}
  {https://doi.org/10.1021/acs.nanolett.2c03069} \BibitemShut {NoStop}%
\bibitem [{\citenamefont {Bucher}\ \emph {et~al.}(2020)\citenamefont {Bucher},
  \citenamefont {Glenn}, \citenamefont {Park}, \citenamefont {Lukin},\ and\
  \citenamefont {Walsworth}}]{PhysRevX.10.021053}%
  \BibitemOpen
  \bibfield  {author} {\bibinfo {author} {\bibfnamefont {D.~B.}\ \bibnamefont
  {Bucher}}, \bibinfo {author} {\bibfnamefont {D.~R.}\ \bibnamefont {Glenn}},
  \bibinfo {author} {\bibfnamefont {H.}~\bibnamefont {Park}}, \bibinfo {author}
  {\bibfnamefont {M.~D.}\ \bibnamefont {Lukin}}, \ and\ \bibinfo {author}
  {\bibfnamefont {R.~L.}\ \bibnamefont {Walsworth}},\ }\bibfield  {title}
  {\enquote {\bibinfo {title} {Hyperpolarization-enhanced nmr spectroscopy with
  femtomole sensitivity using quantum defects in diamond},}\ }\href {\doibase
  10.1103/PhysRevX.10.021053} {\bibfield  {journal} {\bibinfo  {journal} {Phys.
  Rev. X}\ }\textbf {\bibinfo {volume} {10}},\ \bibinfo {pages} {021053}
  (\bibinfo {year} {2020})}\BibitemShut {NoStop}%
\bibitem [{\citenamefont {Wilson}\ \emph {et~al.}(2009)\citenamefont {Wilson},
  \citenamefont {Hurd}, \citenamefont {Keshari}, \citenamefont {Criekinge},
  \citenamefont {Chen}, \citenamefont {Nelson}, \citenamefont {Vigneron},\ and\
  \citenamefont {Kurhanewicz}}]{doi:10.1073/pnas.0810190106}%
  \BibitemOpen
  \bibfield  {author} {\bibinfo {author} {\bibfnamefont {D.~M.}\ \bibnamefont
  {Wilson}}, \bibinfo {author} {\bibfnamefont {R.~E.}\ \bibnamefont {Hurd}},
  \bibinfo {author} {\bibfnamefont {K.}~\bibnamefont {Keshari}}, \bibinfo
  {author} {\bibfnamefont {M.~V.}\ \bibnamefont {Criekinge}}, \bibinfo {author}
  {\bibfnamefont {A.~P.}\ \bibnamefont {Chen}}, \bibinfo {author}
  {\bibfnamefont {S.~J.}\ \bibnamefont {Nelson}}, \bibinfo {author}
  {\bibfnamefont {D.~B.}\ \bibnamefont {Vigneron}}, \ and\ \bibinfo {author}
  {\bibfnamefont {J.}~\bibnamefont {Kurhanewicz}},\ }\bibfield  {title}
  {\enquote {\bibinfo {title} {Generation of hyperpolarized substrates by
  secondary labeling with [1,1-<sup>13</sup>c] acetic anhydride},}\ }\href
  {\doibase 10.1073/pnas.0810190106} {\bibfield  {journal} {\bibinfo  {journal}
  {Proceedings of the National Academy of Sciences}\ }\textbf {\bibinfo
  {volume} {106}},\ \bibinfo {pages} {5503--5507} (\bibinfo {year} {2009})},\
  \Eprint
  {http://arxiv.org/abs/https://www.pnas.org/doi/pdf/10.1073/pnas.0810190106}
  {https://www.pnas.org/doi/pdf/10.1073/pnas.0810190106} \BibitemShut {NoStop}%
\bibitem [{\citenamefont {Marshall}\ \emph {et~al.}(2023)\citenamefont
  {Marshall}, \citenamefont {Salhov}, \citenamefont {Gierse}, \citenamefont
  {Müller}, \citenamefont {Keim}, \citenamefont {Lucas}, \citenamefont
  {Parker}, \citenamefont {Scheuer}, \citenamefont {Vassiliou}, \citenamefont
  {Neumann}, \citenamefont {Jelezko}, \citenamefont {Retzker}, \citenamefont
  {Blanchard}, \citenamefont {Schwartz},\ and\ \citenamefont
  {Knecht}}]{doi:10.1021/acs.jpclett.2c03785}%
  \BibitemOpen
  \bibfield  {author} {\bibinfo {author} {\bibfnamefont {A.}~\bibnamefont
  {Marshall}}, \bibinfo {author} {\bibfnamefont {A.}~\bibnamefont {Salhov}},
  \bibinfo {author} {\bibfnamefont {M.}~\bibnamefont {Gierse}}, \bibinfo
  {author} {\bibfnamefont {C.}~\bibnamefont {Müller}}, \bibinfo {author}
  {\bibfnamefont {M.}~\bibnamefont {Keim}}, \bibinfo {author} {\bibfnamefont
  {S.}~\bibnamefont {Lucas}}, \bibinfo {author} {\bibfnamefont
  {A.}~\bibnamefont {Parker}}, \bibinfo {author} {\bibfnamefont
  {J.}~\bibnamefont {Scheuer}}, \bibinfo {author} {\bibfnamefont
  {C.}~\bibnamefont {Vassiliou}}, \bibinfo {author} {\bibfnamefont
  {P.}~\bibnamefont {Neumann}}, \bibinfo {author} {\bibfnamefont
  {F.}~\bibnamefont {Jelezko}}, \bibinfo {author} {\bibfnamefont
  {A.}~\bibnamefont {Retzker}}, \bibinfo {author} {\bibfnamefont {J.~W.}\
  \bibnamefont {Blanchard}}, \bibinfo {author} {\bibfnamefont {I.}~\bibnamefont
  {Schwartz}}, \ and\ \bibinfo {author} {\bibfnamefont {S.}~\bibnamefont
  {Knecht}},\ }\bibfield  {title} {\enquote {\bibinfo {title} {Radio-frequency
  sweeps at microtesla fields for parahydrogen-induced polarization of
  biomolecules},}\ }\href {\doibase 10.1021/acs.jpclett.2c03785} {\bibfield
  {journal} {\bibinfo  {journal} {The Journal of Physical Chemistry Letters}\
  }\textbf {\bibinfo {volume} {14}},\ \bibinfo {pages} {2125--2132} (\bibinfo
  {year} {2023})},\ \bibinfo {note} {pMID: 36802642},\ \Eprint
  {http://arxiv.org/abs/https://doi.org/10.1021/acs.jpclett.2c03785}
  {https://doi.org/10.1021/acs.jpclett.2c03785} \BibitemShut {NoStop}%
\bibitem [{\citenamefont {Marshall}\ \emph {et~al.}(2022)\citenamefont
  {Marshall}, \citenamefont {Reisser}, \citenamefont {Rembold}, \citenamefont
  {M\"uller}, \citenamefont {Scheuer}, \citenamefont {Gierse}, \citenamefont
  {Eichhorn}, \citenamefont {Steiner}, \citenamefont {Hautle}, \citenamefont
  {Calarco}, \citenamefont {Jelezko}, \citenamefont {Plenio}, \citenamefont
  {Montangero}, \citenamefont {Schwartz}, \citenamefont {M\"uller},\ and\
  \citenamefont {Neumann}}]{PhysRevResearch.4.043179}%
  \BibitemOpen
  \bibfield  {author} {\bibinfo {author} {\bibfnamefont {A.}~\bibnamefont
  {Marshall}}, \bibinfo {author} {\bibfnamefont {T.}~\bibnamefont {Reisser}},
  \bibinfo {author} {\bibfnamefont {P.}~\bibnamefont {Rembold}}, \bibinfo
  {author} {\bibfnamefont {C.}~\bibnamefont {M\"uller}}, \bibinfo {author}
  {\bibfnamefont {J.}~\bibnamefont {Scheuer}}, \bibinfo {author} {\bibfnamefont
  {M.}~\bibnamefont {Gierse}}, \bibinfo {author} {\bibfnamefont
  {T.}~\bibnamefont {Eichhorn}}, \bibinfo {author} {\bibfnamefont {J.~M.}\
  \bibnamefont {Steiner}}, \bibinfo {author} {\bibfnamefont {P.}~\bibnamefont
  {Hautle}}, \bibinfo {author} {\bibfnamefont {T.}~\bibnamefont {Calarco}},
  \bibinfo {author} {\bibfnamefont {F.}~\bibnamefont {Jelezko}}, \bibinfo
  {author} {\bibfnamefont {M.~B.}\ \bibnamefont {Plenio}}, \bibinfo {author}
  {\bibfnamefont {S.}~\bibnamefont {Montangero}}, \bibinfo {author}
  {\bibfnamefont {I.}~\bibnamefont {Schwartz}}, \bibinfo {author}
  {\bibfnamefont {M.~M.}\ \bibnamefont {M\"uller}}, \ and\ \bibinfo {author}
  {\bibfnamefont {P.}~\bibnamefont {Neumann}},\ }\bibfield  {title} {\enquote
  {\bibinfo {title} {Macroscopic hyperpolarization enhanced with quantum
  optimal control},}\ }\href {\doibase 10.1103/PhysRevResearch.4.043179}
  {\bibfield  {journal} {\bibinfo  {journal} {Phys. Rev. Res.}\ }\textbf
  {\bibinfo {volume} {4}},\ \bibinfo {pages} {043179} (\bibinfo {year}
  {2022})}\BibitemShut {NoStop}%
\bibitem [{\citenamefont {Tetienne}\ \emph {et~al.}(2021)\citenamefont
  {Tetienne}, \citenamefont {Hall}, \citenamefont {Healey}, \citenamefont
  {White}, \citenamefont {Sani}, \citenamefont {Separovic},\ and\ \citenamefont
  {Hollenberg}}]{PhysRevB.103.014434}%
  \BibitemOpen
  \bibfield  {author} {\bibinfo {author} {\bibfnamefont {J.-P.}\ \bibnamefont
  {Tetienne}}, \bibinfo {author} {\bibfnamefont {L.~T.}\ \bibnamefont {Hall}},
  \bibinfo {author} {\bibfnamefont {A.~J.}\ \bibnamefont {Healey}}, \bibinfo
  {author} {\bibfnamefont {G.~A.~L.}\ \bibnamefont {White}}, \bibinfo {author}
  {\bibfnamefont {M.-A.}\ \bibnamefont {Sani}}, \bibinfo {author}
  {\bibfnamefont {F.}~\bibnamefont {Separovic}}, \ and\ \bibinfo {author}
  {\bibfnamefont {L.~C.~L.}\ \bibnamefont {Hollenberg}},\ }\bibfield  {title}
  {\enquote {\bibinfo {title} {Prospects for nuclear spin hyperpolarization of
  molecular samples using nitrogen-vacancy centers in diamond},}\ }\href
  {\doibase 10.1103/PhysRevB.103.014434} {\bibfield  {journal} {\bibinfo
  {journal} {Phys. Rev. B}\ }\textbf {\bibinfo {volume} {103}},\ \bibinfo
  {pages} {014434} (\bibinfo {year} {2021})}\BibitemShut {NoStop}%
\bibitem [{\citenamefont {Picollo}\ \emph {et~al.}(2017)\citenamefont
  {Picollo}, \citenamefont {Battiato}, \citenamefont {Boarino}, \citenamefont
  {{Ditalia Tchernij}}, \citenamefont {Enrico}, \citenamefont {Forneris},
  \citenamefont {Gilardino}, \citenamefont {Jakšić}, \citenamefont {Sardi},
  \citenamefont {Skukan}, \citenamefont {Tengattini}, \citenamefont {Olivero},
  \citenamefont {Re},\ and\ \citenamefont {Vittone}}]{PICOLLO2017193}%
  \BibitemOpen
  \bibfield  {author} {\bibinfo {author} {\bibfnamefont {F.}~\bibnamefont
  {Picollo}}, \bibinfo {author} {\bibfnamefont {A.}~\bibnamefont {Battiato}},
  \bibinfo {author} {\bibfnamefont {L.}~\bibnamefont {Boarino}}, \bibinfo
  {author} {\bibfnamefont {S.}~\bibnamefont {{Ditalia Tchernij}}}, \bibinfo
  {author} {\bibfnamefont {E.}~\bibnamefont {Enrico}}, \bibinfo {author}
  {\bibfnamefont {J.}~\bibnamefont {Forneris}}, \bibinfo {author}
  {\bibfnamefont {A.}~\bibnamefont {Gilardino}}, \bibinfo {author}
  {\bibfnamefont {M.}~\bibnamefont {Jakšić}}, \bibinfo {author}
  {\bibfnamefont {F.}~\bibnamefont {Sardi}}, \bibinfo {author} {\bibfnamefont
  {N.}~\bibnamefont {Skukan}}, \bibinfo {author} {\bibfnamefont
  {A.}~\bibnamefont {Tengattini}}, \bibinfo {author} {\bibfnamefont
  {P.}~\bibnamefont {Olivero}}, \bibinfo {author} {\bibfnamefont
  {A.}~\bibnamefont {Re}}, \ and\ \bibinfo {author} {\bibfnamefont
  {E.}~\bibnamefont {Vittone}},\ }\bibfield  {title} {\enquote {\bibinfo
  {title} {Fabrication of monolithic microfluidic channels in diamond with ion
  beam lithography},}\ }\href {\doibase
  https://doi.org/10.1016/j.nimb.2017.01.062} {\bibfield  {journal} {\bibinfo
  {journal} {Nuclear Instruments and Methods in Physics Research Section B:
  Beam Interactions with Materials and Atoms}\ }\textbf {\bibinfo {volume}
  {404}},\ \bibinfo {pages} {193--197} (\bibinfo {year} {2017})},\ \bibinfo
  {note} {proceedings of the 15th International Conference on Nuclear
  Microprobe Technology and Applications}\BibitemShut {NoStop}%
\bibitem [{\citenamefont {Eaton}\ \emph {et~al.}(2019)\citenamefont {Eaton},
  \citenamefont {Hadden}, \citenamefont {Bharadwaj}, \citenamefont {Forneris},
  \citenamefont {Picollo}, \citenamefont {Bosia}, \citenamefont {Sotillo},
  \citenamefont {Giakoumaki}, \citenamefont {Jedrkiewicz}, \citenamefont
  {Chiappini}, \citenamefont {Ferrari}, \citenamefont {Osellame}, \citenamefont
  {Barclay}, \citenamefont {Olivero},\ and\ \citenamefont
  {Ramponi}}]{https://doi.org/10.1002/qute.201900006}%
  \BibitemOpen
  \bibfield  {author} {\bibinfo {author} {\bibfnamefont {S.~M.}\ \bibnamefont
  {Eaton}}, \bibinfo {author} {\bibfnamefont {J.~P.}\ \bibnamefont {Hadden}},
  \bibinfo {author} {\bibfnamefont {V.}~\bibnamefont {Bharadwaj}}, \bibinfo
  {author} {\bibfnamefont {J.}~\bibnamefont {Forneris}}, \bibinfo {author}
  {\bibfnamefont {F.}~\bibnamefont {Picollo}}, \bibinfo {author} {\bibfnamefont
  {F.}~\bibnamefont {Bosia}}, \bibinfo {author} {\bibfnamefont
  {B.}~\bibnamefont {Sotillo}}, \bibinfo {author} {\bibfnamefont {A.~N.}\
  \bibnamefont {Giakoumaki}}, \bibinfo {author} {\bibfnamefont
  {O.}~\bibnamefont {Jedrkiewicz}}, \bibinfo {author} {\bibfnamefont
  {A.}~\bibnamefont {Chiappini}}, \bibinfo {author} {\bibfnamefont
  {M.}~\bibnamefont {Ferrari}}, \bibinfo {author} {\bibfnamefont
  {R.}~\bibnamefont {Osellame}}, \bibinfo {author} {\bibfnamefont {P.~E.}\
  \bibnamefont {Barclay}}, \bibinfo {author} {\bibfnamefont {P.}~\bibnamefont
  {Olivero}}, \ and\ \bibinfo {author} {\bibfnamefont {R.}~\bibnamefont
  {Ramponi}},\ }\bibfield  {title} {\enquote {\bibinfo {title} {Quantum
  micro–nano devices fabricated in diamond by femtosecond laser and ion
  irradiation},}\ }\href {\doibase https://doi.org/10.1002/qute.201900006}
  {\bibfield  {journal} {\bibinfo  {journal} {Advanced Quantum Technologies}\
  }\textbf {\bibinfo {volume} {2}},\ \bibinfo {pages} {1900006} (\bibinfo
  {year} {2019})},\ \Eprint
  {http://arxiv.org/abs/https://onlinelibrary.wiley.com/doi/pdf/10.1002/qute.201900006}
  {https://onlinelibrary.wiley.com/doi/pdf/10.1002/qute.201900006} \BibitemShut
  {NoStop}%
\bibitem [{\citenamefont {Jedrkiewicz}\ \emph {et~al.}(2017)\citenamefont
  {Jedrkiewicz}, \citenamefont {Kumar}, \citenamefont {Sotillo}, \citenamefont
  {Bollani}, \citenamefont {Chiappini}, \citenamefont {Ferrari}, \citenamefont
  {Ramponi}, \citenamefont {Trapani},\ and\ \citenamefont
  {Eaton}}]{Jedrkiewicz:17}%
  \BibitemOpen
  \bibfield  {author} {\bibinfo {author} {\bibfnamefont {O.}~\bibnamefont
  {Jedrkiewicz}}, \bibinfo {author} {\bibfnamefont {S.}~\bibnamefont {Kumar}},
  \bibinfo {author} {\bibfnamefont {B.}~\bibnamefont {Sotillo}}, \bibinfo
  {author} {\bibfnamefont {M.}~\bibnamefont {Bollani}}, \bibinfo {author}
  {\bibfnamefont {A.}~\bibnamefont {Chiappini}}, \bibinfo {author}
  {\bibfnamefont {M.}~\bibnamefont {Ferrari}}, \bibinfo {author} {\bibfnamefont
  {R.}~\bibnamefont {Ramponi}}, \bibinfo {author} {\bibfnamefont {P.~D.}\
  \bibnamefont {Trapani}}, \ and\ \bibinfo {author} {\bibfnamefont {S.~M.}\
  \bibnamefont {Eaton}},\ }\bibfield  {title} {\enquote {\bibinfo {title}
  {Pulsed bessel beam-induced microchannels on a diamond surface for versatile
  microfluidic and sensing applications},}\ }\href {\doibase
  10.1364/OME.7.001962} {\bibfield  {journal} {\bibinfo  {journal} {Opt. Mater.
  Express}\ }\textbf {\bibinfo {volume} {7}},\ \bibinfo {pages} {1962--1970}
  (\bibinfo {year} {2017})}\BibitemShut {NoStop}%
\bibitem [{\citenamefont {Pin}, \citenamefont {Otsuka},\ and\ \citenamefont
  {Sasaki}(2020)}]{doi:10.1021/acsanm.0c00274}%
  \BibitemOpen
  \bibfield  {author} {\bibinfo {author} {\bibfnamefont {C.}~\bibnamefont
  {Pin}}, \bibinfo {author} {\bibfnamefont {R.}~\bibnamefont {Otsuka}}, \ and\
  \bibinfo {author} {\bibfnamefont {K.}~\bibnamefont {Sasaki}},\ }\bibfield
  {title} {\enquote {\bibinfo {title} {Optical transport and sorting of
  fluorescent nanodiamonds inside a tapered glass capillary: Optical sorting of
  nanomaterials at the femtonewton scale},}\ }\href {\doibase
  10.1021/acsanm.0c00274} {\bibfield  {journal} {\bibinfo  {journal} {ACS
  Applied Nano Materials}\ }\textbf {\bibinfo {volume} {3}},\ \bibinfo {pages}
  {4127--4134} (\bibinfo {year} {2020})},\ \Eprint
  {http://arxiv.org/abs/https://doi.org/10.1021/acsanm.0c00274}
  {https://doi.org/10.1021/acsanm.0c00274} \BibitemShut {NoStop}%
\bibitem [{\citenamefont {Dr{\'e}au}\ \emph {et~al.}(2011)\citenamefont
  {Dr{\'e}au}, \citenamefont {Lesik}, \citenamefont {Rondin}, \citenamefont
  {Spinicelli}, \citenamefont {Arcizet}, \citenamefont {Roch},\ and\
  \citenamefont {Jacques}}]{dreau2011avoiding}%
  \BibitemOpen
  \bibfield  {author} {\bibinfo {author} {\bibfnamefont {A.}~\bibnamefont
  {Dr{\'e}au}}, \bibinfo {author} {\bibfnamefont {M.}~\bibnamefont {Lesik}},
  \bibinfo {author} {\bibfnamefont {L.}~\bibnamefont {Rondin}}, \bibinfo
  {author} {\bibfnamefont {P.}~\bibnamefont {Spinicelli}}, \bibinfo {author}
  {\bibfnamefont {O.}~\bibnamefont {Arcizet}}, \bibinfo {author} {\bibfnamefont
  {J.-F.}\ \bibnamefont {Roch}}, \ and\ \bibinfo {author} {\bibfnamefont
  {V.}~\bibnamefont {Jacques}},\ }\bibfield  {title} {\enquote {\bibinfo
  {title} {Avoiding power broadening in optically detected magnetic resonance
  of single nv defects for enhanced dc magnetic field sensitivity},}\
  }\href@noop {} {\bibfield  {journal} {\bibinfo  {journal} {Phys. Rev. B}\
  }\textbf {\bibinfo {volume} {84}},\ \bibinfo {pages} {195204} (\bibinfo
  {year} {2011})}\BibitemShut {NoStop}%
\bibitem [{\citenamefont {Clevenson}\ \emph {et~al.}(2015)\citenamefont
  {Clevenson}, \citenamefont {Trusheim}, \citenamefont {Teale}, \citenamefont
  {Schr{\"o}der}, \citenamefont {Braje},\ and\ \citenamefont
  {Englund}}]{clevenson2015broadband}%
  \BibitemOpen
  \bibfield  {author} {\bibinfo {author} {\bibfnamefont {H.}~\bibnamefont
  {Clevenson}}, \bibinfo {author} {\bibfnamefont {M.~E.}\ \bibnamefont
  {Trusheim}}, \bibinfo {author} {\bibfnamefont {C.}~\bibnamefont {Teale}},
  \bibinfo {author} {\bibfnamefont {T.}~\bibnamefont {Schr{\"o}der}}, \bibinfo
  {author} {\bibfnamefont {D.}~\bibnamefont {Braje}}, \ and\ \bibinfo {author}
  {\bibfnamefont {D.}~\bibnamefont {Englund}},\ }\bibfield  {title} {\enquote
  {\bibinfo {title} {Broadband magnetometry and temperature sensing with a
  light-trapping diamond waveguide},}\ }\href@noop {} {\bibfield  {journal}
  {\bibinfo  {journal} {Nature Physics}\ }\textbf {\bibinfo {volume} {11}},\
  \bibinfo {pages} {393--397} (\bibinfo {year} {2015})}\BibitemShut {NoStop}%
\bibitem [{\citenamefont {Chen}\ \emph {et~al.}(2017)\citenamefont {Chen},
  \citenamefont {Salter}, \citenamefont {Knauer}, \citenamefont {Weng},
  \citenamefont {Frangeskou}, \citenamefont {Stephen}, \citenamefont {Ishmael},
  \citenamefont {Dolan}, \citenamefont {Johnson}, \citenamefont {Green} \emph
  {et~al.}}]{chen2017laser}%
  \BibitemOpen
  \bibfield  {author} {\bibinfo {author} {\bibfnamefont {Y.-C.}\ \bibnamefont
  {Chen}}, \bibinfo {author} {\bibfnamefont {P.~S.}\ \bibnamefont {Salter}},
  \bibinfo {author} {\bibfnamefont {S.}~\bibnamefont {Knauer}}, \bibinfo
  {author} {\bibfnamefont {L.}~\bibnamefont {Weng}}, \bibinfo {author}
  {\bibfnamefont {A.~C.}\ \bibnamefont {Frangeskou}}, \bibinfo {author}
  {\bibfnamefont {C.~J.}\ \bibnamefont {Stephen}}, \bibinfo {author}
  {\bibfnamefont {S.~N.}\ \bibnamefont {Ishmael}}, \bibinfo {author}
  {\bibfnamefont {P.~R.}\ \bibnamefont {Dolan}}, \bibinfo {author}
  {\bibfnamefont {S.}~\bibnamefont {Johnson}}, \bibinfo {author} {\bibfnamefont
  {B.~L.}\ \bibnamefont {Green}},  \emph {et~al.},\ }\bibfield  {title}
  {\enquote {\bibinfo {title} {Laser writing of coherent colour centres in
  diamond},}\ }\href@noop {} {\bibfield  {journal} {\bibinfo  {journal} {Nature
  Photonics}\ }\textbf {\bibinfo {volume} {11}},\ \bibinfo {pages} {77--80}
  (\bibinfo {year} {2017})}\BibitemShut {NoStop}%
\bibitem [{\citenamefont {Hatano}\ \emph {et~al.}(2021)\citenamefont {Hatano},
  \citenamefont {Shin}, \citenamefont {Nishitani}, \citenamefont {Iwatsuka},
  \citenamefont {Masuyama}, \citenamefont {Sugiyama}, \citenamefont {Ishii},
  \citenamefont {Onoda}, \citenamefont {Ohshima}, \citenamefont {Arai},
  \citenamefont {Iwasaki},\ and\ \citenamefont {Hatano}}]{10.1063/5.0031502}%
  \BibitemOpen
  \bibfield  {author} {\bibinfo {author} {\bibfnamefont {Y.}~\bibnamefont
  {Hatano}}, \bibinfo {author} {\bibfnamefont {J.}~\bibnamefont {Shin}},
  \bibinfo {author} {\bibfnamefont {D.}~\bibnamefont {Nishitani}}, \bibinfo
  {author} {\bibfnamefont {H.}~\bibnamefont {Iwatsuka}}, \bibinfo {author}
  {\bibfnamefont {Y.}~\bibnamefont {Masuyama}}, \bibinfo {author}
  {\bibfnamefont {H.}~\bibnamefont {Sugiyama}}, \bibinfo {author}
  {\bibfnamefont {M.}~\bibnamefont {Ishii}}, \bibinfo {author} {\bibfnamefont
  {S.}~\bibnamefont {Onoda}}, \bibinfo {author} {\bibfnamefont
  {T.}~\bibnamefont {Ohshima}}, \bibinfo {author} {\bibfnamefont
  {K.}~\bibnamefont {Arai}}, \bibinfo {author} {\bibfnamefont {T.}~\bibnamefont
  {Iwasaki}}, \ and\ \bibinfo {author} {\bibfnamefont {M.}~\bibnamefont
  {Hatano}},\ }\bibfield  {title} {\enquote {\bibinfo {title} {{Simultaneous
  thermometry and magnetometry using a fiber-coupled quantum diamond
  sensor}},}\ }\href {\doibase 10.1063/5.0031502} {\bibfield  {journal}
  {\bibinfo  {journal} {Applied Physics Letters}\ }\textbf {\bibinfo {volume}
  {118}},\ \bibinfo {pages} {034001} (\bibinfo {year} {2021})},\ \Eprint
  {http://arxiv.org/abs/https://pubs.aip.org/aip/apl/article-pdf/doi/10.1063/5.0031502/14543352/034001\_1\_online.pdf}
  {https://pubs.aip.org/aip/apl/article-pdf/doi/10.1063/5.0031502/14543352/034001\_1\_online.pdf}
  \BibitemShut {NoStop}%
\bibitem [{\citenamefont {Shim}\ \emph {et~al.}(2022)\citenamefont {Shim},
  \citenamefont {Lee}, \citenamefont {Ghimire}, \citenamefont {Hwang},
  \citenamefont {Lee}, \citenamefont {Kim}, \citenamefont {Turner},
  \citenamefont {Hart}, \citenamefont {Walsworth},\ and\ \citenamefont
  {Oh}}]{PhysRevApplied.17.014009}%
  \BibitemOpen
  \bibfield  {author} {\bibinfo {author} {\bibfnamefont {J.~H.}\ \bibnamefont
  {Shim}}, \bibinfo {author} {\bibfnamefont {S.-J.}\ \bibnamefont {Lee}},
  \bibinfo {author} {\bibfnamefont {S.}~\bibnamefont {Ghimire}}, \bibinfo
  {author} {\bibfnamefont {J.~I.}\ \bibnamefont {Hwang}}, \bibinfo {author}
  {\bibfnamefont {K.-G.}\ \bibnamefont {Lee}}, \bibinfo {author} {\bibfnamefont
  {K.}~\bibnamefont {Kim}}, \bibinfo {author} {\bibfnamefont {M.~J.}\
  \bibnamefont {Turner}}, \bibinfo {author} {\bibfnamefont {C.~A.}\
  \bibnamefont {Hart}}, \bibinfo {author} {\bibfnamefont {R.~L.}\ \bibnamefont
  {Walsworth}}, \ and\ \bibinfo {author} {\bibfnamefont {S.}~\bibnamefont
  {Oh}},\ }\bibfield  {title} {\enquote {\bibinfo {title} {Multiplexed sensing
  of magnetic field and temperature in real time using a nitrogen-vacancy
  ensemble in diamond},}\ }\href {\doibase 10.1103/PhysRevApplied.17.014009}
  {\bibfield  {journal} {\bibinfo  {journal} {Phys. Rev. Appl.}\ }\textbf
  {\bibinfo {volume} {17}},\ \bibinfo {pages} {014009} (\bibinfo {year}
  {2022})}\BibitemShut {NoStop}%
\bibitem [{\citenamefont {Hart}\ and\ \citenamefont
  {Knowles}(2023)}]{10.3389/frqst.2023.1220015}%
  \BibitemOpen
  \bibfield  {author} {\bibinfo {author} {\bibfnamefont {J.~W.}\ \bibnamefont
  {Hart}}\ and\ \bibinfo {author} {\bibfnamefont {H.~S.}\ \bibnamefont
  {Knowles}},\ }\bibfield  {title} {\enquote {\bibinfo {title} {Multimodal
  quantum metrology in living systems using nitrogen-vacancy centres in diamond
  nanocrystals},}\ }\href {\doibase 10.3389/frqst.2023.1220015} {\bibfield
  {journal} {\bibinfo  {journal} {Frontiers in Quantum Science and Technology}\
  }\textbf {\bibinfo {volume} {2}} (\bibinfo {year} {2023}),\
  10.3389/frqst.2023.1220015}\BibitemShut {NoStop}%
\bibitem [{\citenamefont {Castelletto}\ and\ \citenamefont
  {Boretti}(2020)}]{castelletto2020silicon}%
  \BibitemOpen
  \bibfield  {author} {\bibinfo {author} {\bibfnamefont {S.}~\bibnamefont
  {Castelletto}}\ and\ \bibinfo {author} {\bibfnamefont {A.}~\bibnamefont
  {Boretti}},\ }\bibfield  {title} {\enquote {\bibinfo {title} {Silicon carbide
  color centers for quantum applications},}\ }\href@noop {} {\bibfield
  {journal} {\bibinfo  {journal} {Journal of Physics: Photonics}\ }\textbf
  {\bibinfo {volume} {2}},\ \bibinfo {pages} {022001} (\bibinfo {year}
  {2020})}\BibitemShut {NoStop}%
\bibitem [{\citenamefont {Jiang}\ \emph {et~al.}(2023)\citenamefont {Jiang},
  \citenamefont {Cai}, \citenamefont {Cernansky}, \citenamefont {Liu},\ and\
  \citenamefont {Gao}}]{doi:10.1126/sciadv.adg2080}%
  \BibitemOpen
  \bibfield  {author} {\bibinfo {author} {\bibfnamefont {Z.}~\bibnamefont
  {Jiang}}, \bibinfo {author} {\bibfnamefont {H.}~\bibnamefont {Cai}}, \bibinfo
  {author} {\bibfnamefont {R.}~\bibnamefont {Cernansky}}, \bibinfo {author}
  {\bibfnamefont {X.}~\bibnamefont {Liu}}, \ and\ \bibinfo {author}
  {\bibfnamefont {W.}~\bibnamefont {Gao}},\ }\bibfield  {title} {\enquote
  {\bibinfo {title} {Quantum sensing of radio-frequency signal with nv centers
  in sic},}\ }\href {\doibase 10.1126/sciadv.adg2080} {\bibfield  {journal}
  {\bibinfo  {journal} {Science Advances}\ }\textbf {\bibinfo {volume} {9}},\
  \bibinfo {pages} {eadg2080} (\bibinfo {year} {2023})},\ \Eprint
  {http://arxiv.org/abs/https://www.science.org/doi/pdf/10.1126/sciadv.adg2080}
  {https://www.science.org/doi/pdf/10.1126/sciadv.adg2080} \BibitemShut
  {NoStop}%
\bibitem [{\citenamefont {Gottscholl}\ \emph {et~al.}(2021)\citenamefont
  {Gottscholl}, \citenamefont {Diez}, \citenamefont {Soltamov}, \citenamefont
  {Kasper}, \citenamefont {Krau{\ss}e}, \citenamefont {Sperlich}, \citenamefont
  {Kianinia}, \citenamefont {Bradac}, \citenamefont {Aharonovich},\ and\
  \citenamefont {Dyakonov}}]{gottscholl2021spin}%
  \BibitemOpen
  \bibfield  {author} {\bibinfo {author} {\bibfnamefont {A.}~\bibnamefont
  {Gottscholl}}, \bibinfo {author} {\bibfnamefont {M.}~\bibnamefont {Diez}},
  \bibinfo {author} {\bibfnamefont {V.}~\bibnamefont {Soltamov}}, \bibinfo
  {author} {\bibfnamefont {C.}~\bibnamefont {Kasper}}, \bibinfo {author}
  {\bibfnamefont {D.}~\bibnamefont {Krau{\ss}e}}, \bibinfo {author}
  {\bibfnamefont {A.}~\bibnamefont {Sperlich}}, \bibinfo {author}
  {\bibfnamefont {M.}~\bibnamefont {Kianinia}}, \bibinfo {author}
  {\bibfnamefont {C.}~\bibnamefont {Bradac}}, \bibinfo {author} {\bibfnamefont
  {I.}~\bibnamefont {Aharonovich}}, \ and\ \bibinfo {author} {\bibfnamefont
  {V.}~\bibnamefont {Dyakonov}},\ }\bibfield  {title} {\enquote {\bibinfo
  {title} {Spin defects in hbn as promising temperature, pressure and magnetic
  field quantum sensors},}\ }\href@noop {} {\bibfield  {journal} {\bibinfo
  {journal} {Nature Communications}\ }\textbf {\bibinfo {volume} {12}},\
  \bibinfo {pages} {4480} (\bibinfo {year} {2021})}\BibitemShut {NoStop}%
\bibitem [{\citenamefont {Vaidya}\ \emph {et~al.}(2023)\citenamefont {Vaidya},
  \citenamefont {Gao}, \citenamefont {Dikshit}, \citenamefont {Aharonovich},\
  and\ \citenamefont {Li}}]{doi:10.1080/23746149.2023.2206049}%
  \BibitemOpen
  \bibfield  {author} {\bibinfo {author} {\bibfnamefont {S.}~\bibnamefont
  {Vaidya}}, \bibinfo {author} {\bibfnamefont {X.}~\bibnamefont {Gao}},
  \bibinfo {author} {\bibfnamefont {S.}~\bibnamefont {Dikshit}}, \bibinfo
  {author} {\bibfnamefont {I.}~\bibnamefont {Aharonovich}}, \ and\ \bibinfo
  {author} {\bibfnamefont {T.}~\bibnamefont {Li}},\ }\bibfield  {title}
  {\enquote {\bibinfo {title} {Quantum sensing and imaging with spin defects in
  hexagonal boron nitride},}\ }\href {\doibase 10.1080/23746149.2023.2206049}
  {\bibfield  {journal} {\bibinfo  {journal} {Advances in Physics: X}\ }\textbf
  {\bibinfo {volume} {8}},\ \bibinfo {pages} {2206049} (\bibinfo {year}
  {2023})},\ \Eprint
  {http://arxiv.org/abs/https://doi.org/10.1080/23746149.2023.2206049}
  {https://doi.org/10.1080/23746149.2023.2206049} \BibitemShut {NoStop}%
\end{thebibliography}

%merlin.mbs aipnum4-1.bst 2010-07-25 4.21a (PWD, AO, DPC) hacked
%Control: key (0)
%Control: author (8) initials jnrlst
%Control: editor formatted (1) identically to author
%Control: production of article title (0) allowed
%Control: page (1) range
%Control: year (1) truncated
%Control: production of eprint (0) enabled
%

\end{document}